\documentclass[10pt, english]{article}
\usepackage{amsmath}
\usepackage{amssymb}
\usepackage{amsthm}

\usepackage[pdftex]{graphicx}
\usepackage{epstopdf}
\graphicspath{{./graphics/}{./matlab_plots/}}

 \newcommand{\bones}{\mathbf{1}}  

 \newcommand{\bzeros}{\mathbf{0}}

\newenvironment{definition}[1][Definition]{\begin{trivlist}
\item[\hskip \labelsep {\bfseries #1}]}{\end{trivlist}}

  \newcommand{\bTheta}{\mbox{\boldmath{$\Theta$}}}
  \newcommand{\bDelta}{\mbox{\boldmath{$\Delta$}}}

  \newcommand{\bvartheta}{\mbox{\boldmath{$\vartheta$}}}

  \newcommand{\bVartheta}{\mbox{\boldmath{$\Upsilon$}}}

  \newcommand{\ba}{\mathbf{a}}

  \newcommand{\bs}{\mathbf{s}}

  \newcommand{\bu}{\mathbf{u}}

  \newcommand{\bw}{\mathbf{w}}

  \newcommand{\bx}{\mathbf{x}}

  \newcommand{\by}{\mathbf{y}}

  \newcommand{\bA}{\mathbf{A}}

  \newcommand{\bD}{\mathbf{D}}

  \newcommand{\bG}{\mathbf{G}}

  \newcommand{\bH}{\mathbf{H}}

  \newcommand{\bI}{\mathbf{I}}

  \newcommand{\bP}{\mathbf{P}}

  \newcommand{\bR}{\mathbf{R}}

  \newcommand{\bS}{\mathbf{S}}

  \newcommand{\bU}{\mathbf{U}}

  \newcommand{\bV}{\mathbf{V}}

  \newcommand{\bW}{\mathbf{W}}

  \newcommand{\bT}{\mathbf{T}}

  \newcommand{\bSigma}{\mathbf{\Sigma}}

\newtheorem{lemma}{Lemma}

\newtheorem{theorem}{Theorem}

\newcommand{\beq}{\begin{equation}}

\newcommand{\enq}{\end{equation}}

\newcommand{\beqa}{\begin{eqnarray}}

\newcommand{\enqa}{\end{eqnarray}}

\newcommand{\bea}{\begin{array}}

\newcommand{\ena}{\end{array}}

\newcommand{\bef}{\begin{figure}}

\newcommand{\enf}{\end{figure}}

\newcommand{\bds}{\begin {itemize}}

\newcommand{\eds}{\end {itemize}}

\newcommand{\bdf}{\begin{definition}}

\newcommand{\blm}{\begin{lemma}}

\newcommand{\edf}{\end{definition}}

\newcommand{\elm}{\end{lemma}}

\newcommand{\bthm}{\begin{theorem}}

\newcommand{\ethm}{\end{theorem}}

\newcommand{\cA}{{\ensuremath{\mathcal{A}}}}

\newcommand{\cF}{{\ensuremath{\mathcal{F}}}}

\newcommand{\cR}{{\ensuremath{\mathcal{R}}}}

\newcommand{\cU}{{\ensuremath{\mathcal{U}}}}

\newcommand{\Ntrx}{N_{\mathrm{trx}}}
\newcommand{\Nt}{N_{\mathrm{t}}}

\newcommand{\thetad}{\theta_{\mathrm{d}}}
\newcommand{\thetadB}{\theta_{3, \, \mathrm{dB}}}
\newcommand{\bthetaSLL}{\mbox{\boldmath{$\theta$}}_{\mathrm{SLL}}}
\newcommand{\bDfb}{\mathbf{D}_{\textrm{fb}}}
\newcommand{\bRfb}{\mathbf{R}_{\textrm{fb}}}

\begin{document}
\title{Hybrid RF and Digital Beamformer for Cellular Networks: Algorithms, Microwave Architectures and Measurements}
\author{Vijay Venkateswaran, Florian Pivit and Lei Guan
\thanks{The authors are with Bell Laboratories, Alcatel Lucent, Dublin, Ireland.}}
\maketitle

\begin{abstract}

Modern wireless communication networks, particularly cellular networks utilize multiple antennas to improve the capacity and signal coverage. In these systems, typically an active transceiver is  connected to each antenna. However, this one-to-one mapping between transceivers and antennas will dramatically increase the cost and complexity of a large phased antenna array system.

In this paper, firstly we propose a \emph{partially adaptive} beamformer architecture where a reduced number of transceivers with a digital beamformer (DBF) is connected to an increased number of antennas through an RF beamforming network (RFBN). Then, based on the proposed architecture, we present a methodology to derive the minimum number of transceivers that are required for marco-cell and small-cell base stations, respectively. Subsequently, in order to achieve optimal beampatterns with given cellular standard requirements and RF operational constraints, we propose efficient algorithms to jointly design DBF and RFBN. Starting from the proposed algorithms, we specify generic microwave RFBNs for optimal marco-cell and small-cell networks. In order to verify the proposed approaches, we compare the performance of RFBN using simulations and anechoic chamber measurements. Experimental measurement results confirm the robustness and performance of the proposed hybrid DBF-RFBN concept eventually ensuring that theoretical multi-antenna capacity and coverage are achieved at a little incremental cost. 

\end{abstract}

\section{Introduction}
\label{sec:1}

In a typical cellular base-station, a passive antenna array is usually connected to an RF transceiver in the form of so-called remote radio head, where each transmitted and received signal is shaped by the same beam. Though this passive architecture is quite simple, it has several disadvantages in terms of its applications in  4G and future/5G wireless communications: a) it does not allow spatial separation of multiple users, which can be considered as a very efficient way to utilize limited frequency spectrum; b) it does not improve the signal to noise ratio (SNR) at the user equipment via the use of advanced beamforming technology, which has been largely accepted as a potential key technology for enabling the 5G wireless communications.

In order to improve spectral efficiency and reduce the interference levels a multi-antenna RF transmitter architecture is being proposed for cellular base stations, where each antenna is connected to a dedicated RF chain and a baseband beamformer \cite{Godara:aa_intro}  as shown in Fig. \ref{fig:1}(a). Such antenna arrays with active RF components are commonly referred as active antenna arrays (AAA). This approach allows us to form multiple beams at the same time to/from the same array. 

\bef
\begin{center}
\includegraphics[width=0.8\columnwidth]{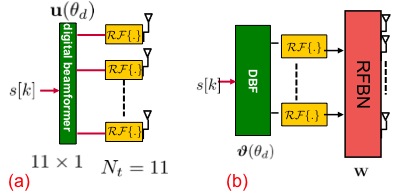}
\end{center}
\caption{\footnotesize{Adaptive antenna array architecture: (a) full dimension active antenna array (AAA) architecture with $\Ntrx=\Nt$ transceivers (denoted by $\cR\cF\{.\}$) and digital beamformer $\bu(\thetad)$ for beamtilt range $\thetad \, \in \cR_{\theta}$ (b) Architecture with adaptive $\bvartheta(\thetad)$ beamforming signals connected to $\Nt \times \Ntrx $ RF beamforming network $\bW$ and finally to $\Nt$ antennas.}}
\label{fig:1}
\enf

\subsection{Setup and Objective} 

A full-size AAA architecture, i.e., each antenna element of an array connected to a dedicated RF chain, significantly increases the cost, weight and overall power consumption, because of its inherent one-to-one mapping between antennas and RF transceivers. In order to reduce the complexity of a full-size AAA architecture, we propose a partially adaptive beamformer based modified AAA architecture, where a digital beamformer (DBF) with a reduced number of RF transceivers is connected to an increased number of antennas through an RF beamforming network (RFBN). This architectural modification imposes a complex set of performance requirements such as spectral mask, microwave/insertion loss, side lobe suppression, effective radiated power, and so on. Thus a comprehensive view to efficiently design RF communication systems is required. 

We consider a setup where an arbitrary $\Ntrx$ transmit signals in digital baseband are converted to RF using a set of $\Ntrx$ RF chains/transceivers. These RF signals are subsequently connected to $\Nt$ antennas using an $\Nt \times \Ntrx$ RFBN as shown in Fig. \ref{fig:1}(b) (where$ \, \Nt  > \Ntrx$).

\subsubsection*{Existing RF beamforming networks} Tunable RF beamformer architectures with reduced number of RF chains have been previously proposed for low-power receivers fully implemented in Silicon \cite{Hajimiri:integrated_phased_arrays, Kautz:phase_shifter, Zarei:phase_shift}. However, the power levels of such  designs are much lower than the power levels required in cellular base-stations, which can easily exceed 47 dBm. At high RF power, the technology to implement RF beamformer is limited to components built in PCBs or suspended strip-line technologies. Some examples of standard passive RF beamformers can be found in \cite{Butler:butler_matrix, Parker:phased_arrays, Quintel:elec_tilt, Mosca:Blass, Tarek:Nolen, Murad:novelButler, Peng:Blass, Chia:Butler}. However, these are non-adaptive designs and a given transmitted signal is shaped by the same beam. Absence of a DBF mean that the coverage and SNR improvement is limited. Additionally, hybrid beamformers have been designed by combining Butler matrices \cite{Butler:butler_matrix} with digital chains for millimeter wave applications \cite{Roh:mmwave}, however the above passive RF beamformer limitations are not addressed.  

As we move towards hybrid networks and consider RF limitations, the hybrid RFBN and DBF must be designed to provide optimal beampattern, minimize microwave loss and reduce complexity. State of the art networks are usually synthesized using empirical (and in some cases systematic) approaches. Hence optimal performance of such architectures might not be guaranteed. In addition, standard RF networks such as Butler matrices \cite{Butler:butler_matrix} are not tailored for a base-station antenna array applications. 

\subsubsection*{Objective} Our aim in this paper is  optimal design of joint RFBN-DBF architectures for a wide variety of cellular architectures. While doing so, we aim to answer some fundamental theoretical and practical questions such as: 
\bds
\item How do we split the RF/analog versus digital functionalities in an adaptive antenna array system? 
\bds
\item i.e. what are the minimum number of digital transceivers and RF chains required for optimal performance?
\eds
\item How do we  achieve optimal main-lobe and sidelobe power level (SLL) for a given RFBN-DBF setup?
\item How do we ensure that the RFBN achieves desired performance with respect to microwave loss, beampattern coverage and recovery in case of transceiver failure? 
-e.g. the only way to account for failures is to use RF combiners. However, in case of dynamic beamforming, the amplitude and phase of transmitted signals do not match at the combiner and in-turn introduces microwave/insertion loss. 
\eds
 
Our design focus varies for various cellular architectures. For example, in a macro-cell setup, the focus is to provide a sharp and narrow vertical beam towards a specific sector, while minimizing the overall loss in the RFBN and satisfying the SLLs. The overall range of beamtilts is usually less ($<20^{\circ}$). On the other hand, in a  small-cell scenario the focus is to provide a set of wide angle beams over an increased range of beamtilts ($\approx 90^{\circ}$). 

\subsection{Signal Processing perspective}
From a signal processing perspective, our proposed setup with RFBN and DBF can be placed in the already known category referred to as 'beamspace processing' \cite{vanVeen:beamforming}. 
In beamspace processing,  signal processing on large arrays (i.e. $\Nt = \Ntrx$) is performed only on a sub-set of transceiver elements instead of on the overall array in order to reduce computational effort. Note that in its traditional sense, \emph{beamspace} processing is entirely done in digital domain. One way to see the RFBN-DBF arrangement is to perform a part of the processing in analog-RF and the rest in digital-baseband. 
 
While beamspace-processing techniques do provide a systematic framework for reduced complexity algorithms, they do not include network setup and SLL constraints in the design specification. In some ways, joint design of RFBN and DBF weights is somewhat similar to \cite{Vijay:apn_journal}, although a complete list of constraints and RF limitations has not been not included in \cite{Vijay:apn_journal} work.

From state of the art passive RF networks, we know how to design a passive phased array system that connects $\Ntrx=1$ transceivers with $\Nt$ antenna elements \cite{Butler:butler_matrix,Parker:phased_arrays}. In this case, all phase and amplitude weights are generated in the RF.  It is also well known how to generate multiple beams with digital beamforming arrays where $\Ntrx=\Nt$ and all phase and amplitude weights are generated in the digital domain \cite{Godara:aa_intro, vanVeen:beamforming} (or in some cases in the analog domain).  The design approaches in both the above cases are straightforward and well documented. 

However, it is not trivial to design RFBN used in combination with DBF to generate multiple beams when $\Ntrx < \Nt$. For example, it is not obvious to ensure optimal beampattern performance for the entire tilt range or cell, while minimizing the microwave loss. In most cases, unique solution cannot be guaranteed. From a signal processing perspective, this work aims to provide a set of rules for identification and placement of feeder network components as well as  routing of digital and RF signals to ensure optimal performance with minimal loss at a reasonable cost.


\subsection{Contributions and Outline}
In this paper, we progressively study various aspects of joint DBF-RFBN design. In Sec. \ref{sec:2}, we specify the antenna array and signal processing model and formulate the design problem. In Sec. \ref{sec:3}, we provide  theoretical bounds on the minimum number of transceivers required for a given objective. Subsequently, we propose algorithms to design the optimal weights of  RFBN and DBF, subject to  performance and design constraints. We recast the joint RFBN-DBF optimization as a convex problem, and use interior point algorithm  \cite{Boyd:cnvx_opt} to find the optimal solution. 

In Sections \ref{sec:4} and \ref{sec:4b}, we represent the RFBN weights using microwave components for macro-cell and small-cell networks. The designs consider macro-cell and small-cell network requirements, and customize the RFBN accordingly. For example, in a high power macro-AAA network, the circuit instantiations must be designed to minimize microwave loss. In this regard, we specify necessary conditions for minimizing microwave loss while factorizing the RFBN into bank of microwave components. On the other hand, in a small-cell network the microwave loss constraint is relaxed in order to provide a wider beamtilt range. 
  
In Sec. \ref{sec:5}, we compare the performance of the proposed architectures and algorithms. In Sec. \ref{sec:6}, we instantiate two RF feeder networks operating at a frequency 2.6 GHz: (a)  Macro-cell AAA RFBN setup  with $\Nt=11 \mbox{ and } \Ntrx=5$ and (b) small-cell AAA RFBN setup $\Nt=6, \mbox{ and }  \Ntrx=3$  and observe their performance. In Sec. \ref{sec:7}, we use the designed setup in anechoic chamber, calibrate the RFBN-DBF setup and perform detailed beampattern  measurements for a macro-cell RFBN with $\Nt=11 \mbox{ and } \Ntrx=5$ followed by conclusions Sec. \ref{sec:8}.

\subsubsection*{Notation}
Lower and upper case bold letters denote vectors and matrices. An over-tilde $(\tilde{.})$ denotes RF signals, while time indexes $(.)$ and $[.]$ respectively denote analog and digital signals. Superscripts $(.)^{T}$, $(.)^{H}$, $(.)^{\dagger}$ and $\|.\|$ respectively denote transpose, Hermitian transpose, pseudo-inverse and Frobenius norm operations. The matrix $\bI_{K}$ denotes an identity matrix and $\odot$ denotes point-wise multiplication of vectors.  $\bzeros_l$ and $\bones_l$ respectively denote $l\times 1$ vectors of zeros and ones.

\section{System model and Proposed Architecture}
\label{sec:2}

\subsection{Data Model}
\label{ssec:2_1}
Consider an $\Nt \times 1$ vector $\tilde{\bx}(t)$ denoting the RF signal  radiated from the antenna array at time $t$. 
In the full dimension AAA setup with $\Nt$ transceivers,  $\tilde{\bx}(t)$ is obtained using an $\Nt \times 1$ DBF vector $\bu(\thetad) = [u_{1}(\thetad), \, \cdots, u_{\Nt}(\thetad)]^{T}$ operating on a data stream $s[k]$ at time  $t=kT$,  followed by $\Nt$ 'RF chains' denoted as $\cR\cF$. (Refer to Fig. \ref{fig:1}(a)). Thus
\[
\tilde{\bx}(t) = \cR\cF\{ \bx[k]\}
\quad \textrm{where} \quad \bx[k] = \bu(\thetad)s[k].\]
Typically, $\bu(\thetad)$ is designed to produce a mainlobe centered at $\thetad$.

Consider our proposed setup with $\Ntrx$ transceivers connected to $\Nt$ radiating elements through a passive RFBN (For example, $\Nt=11$ and $\Ntrx = 5$). Details of RFBN implementation will be explained in Sections \ref{sec:4} - \ref{sec:6}. 
In this case an $\Ntrx\times 1$ DBF vector $\bvartheta(\thetad) = [\vartheta_{1}(\thetad), \, \cdots, \vartheta_{\Ntrx}(\thetad)]^{T}$ operates on the data stream $s[k]$, followed by $\Ntrx$ digital-RF transformation blocks, RFBN matrix $\bW$, and eventually radiated as an $\Nt \times 1$ vector: 
\beq
\tilde{\bx}_{r}(t) = \bW \tilde{\by}(t) 
\label{eq:1_1}\enq
where 
\[
\tilde{\by}(t) = \cR\cF\{\by[k]\} \quad \textrm{and} 
\quad \by[k] = \bvartheta(\thetad)s[k].
\]

We refer to the setup as shown in Fig. \ref{fig:1}(b) as the hybrid beamforming network, since the RFBN is estimated for a specific architecture at the outset and kept fixed. Subsequently, the DBF $\bvartheta(\thetad)$ is adaptively designed for each beamtilt $\thetad$. 

Assuming the antenna elements are equally spaced, the array response $\ba(\theta_i)$ can be modeled as an $\Nt\times1$ vector, which is a function of angle $\theta_{i}$: 
\[
\ba(\theta_{i}) = g(\theta_{i})
\left[ 
\bea{c}
1 \\
e^{j \frac{2 \pi}{\lambda}\delta \cos(\theta_{i})} \\
\vdots \\
e^{j \frac{2 \pi}{\lambda}\delta(\Nt -1) \cos(\theta_{i})}
\ena\right]
\]
where $\delta$ is the spacing between adjacent antennas, $\lambda$ is the wavelength in meters and $g(\theta_i)$ is the antenna characteristic \cite{3GPP:LTE}. Note that the 3GPP transmission standard allows antenna characteristic $g(\theta_{i})$ with a 3-dB beamwidth of either $65^\circ$ or $110^\circ$. These respectively correspond to macro-cell and small-cell antennas in our designs.

\subsection{Full dimension AAA with $\Nt$ transceivers}
\label{ssec:2_2}
The performance of the proposed DBF-RFBN setup is compared with a reference \emph{full dimension AAA setup} having $\Ntrx = \Nt$ transceivers. The performance of the full dimension AAA setup depends on the channel capacity, as well as the adaptive sectorization of the beamformer $\bu(\thetad)$. This performance requirement is specified by the operational constraints and is referred to in this paper as  spectral mask $\bDelta_{\thetad}$. The constraints that make up the spectral mask $\bDelta_{\thetad}$ are explained in detail in Sec. \ref{ssec:2_3} (refer to [C1] - [C6]). In short, the spectral mask includes information regarding the gain and directivity along $\thetad$ as well as the SLLs. 

In full dimension AAA architecture, the objective is to design the adaptive beamformer $\bu(\thetad)$ minimizing the overall mean-squared error:
\beq
\bu_{0} = \arg\min_{\bu(\thetad)} \|\bDelta_{\thetad} - \bA(\theta)\bu(\thetad)\|^2
\label{eq:1_2}
\enq
where 
\[
\bA(\theta) = \left[
\bea{c}
\ba^{T}(-\pi) \\
\vdots \\
\ba^{T}(\pi)]
\ena\right]
\] 
is an $N_\theta \times \Nt$ matrix obtained by stacking the array response vectors. The rows of  $\bA(\theta)$ and length of $\bDelta_{\thetad}$ correspond to the spatial resolution. 
One approach to estimate $\bu(\thetad)$ from (\ref{eq:1_2}) is \[\bu_{0} \approx \bA(\theta)^{\dagger}\bDelta_{\thetad}\] using the least-squares approach \cite{Moon:math_methods}. However, such an approach does not take microwave component design into consideration, would not ensure PAs operating in a linear mode and will not satisfy the required SLLs. Thus the overall design will be sub-optimal from an RF systems perspective.

We will design the full dimensional AAA beamformer weights subject to operational constraints using iterative convex optimization techniques \cite{Boyd:cnvx_opt}. The operational constraints include desired power levels and 3-dB beamwidth along $\thetad$, SLLs and dynamic range of PA output. The optimal full dimensional AAA solution using  convex optimization techniques has been shown in \cite{Palomar:joint_trx_beamforming}, \cite{Yu:convex_beamforming}, so the details of the beamformer design for full dimension AAA are omitted in this paper. 

The full dimensional AAA architecture is not among the main contributions of this paper, however it will serve as our reference design for performance comparisons. 

\subsection{Reduced dimension RFBN Architecture: Problem Formulation}
\label{ssec:2_3}

As mentioned before, our objective is to reduce the number of transceivers and therefore reduce the cost and power consumed by the antenna array. Thus we jointly design the optimal RFBN matrix $\bW$ and DBF vector $\bvartheta(\thetad)$ to satisfy the desired set of spectral mask $\bDelta_{\thetad}$ corresponding to all beamtilts i.e. $ \bDelta_{\thetad}, \, \forall \thetad \in \cR_{\theta} = \{\theta_{1}, \, \cdots, \, \theta_{N_{\theta}}\}$. In this case, $N_{\theta}$ denotes number of sectors. 


Jointly estimating two parameters (such as $\bW$ and $\bvartheta(\thetad)$) for requirements (such as $\bDelta_{\thetad}$) can be solved as a weighted least squares (WLS) problem \cite{Moon:math_methods}. One approach to estimate these parameters is through minimizing the overall mean-square error (MSE): 
\beqa
\{\bW, \, \bvartheta(\thetad)\} 
& = &
\arg\min_{\bW\, \bvartheta(\thetad)} \| \bDelta_{\thetad} - \bA(\theta)\bW\bvartheta(\thetad)\|^2 
\label{eq:1_3} \\
& & \quad \forall \thetad \in \{\theta_{1}, \, \cdots, \, \theta_{N_{\theta}}\} \nonumber
\enqa
In order to minimize the overall cost, while taking practical issues into consideration, the above cost function in (\ref{eq:1_3}) must include the following constraints: 
\begin{itemize}
\item[{[C1]}] The number of transceivers $\Ntrx$ is restricted to a minimum.

\item[{[C2]}] The SLLs are constrained to be at-least 15 dB below the mainlobe. This is to ensure that most of the power is directed towards the desired sector, as well as to limit the interference to neighboring cells/sectors. The 3-dB beamwidth is constrained to be less than $5^\circ$ for a macro-cell and less than $15^\circ$ for a small-cell setup.

\item[{[C3]}] The RFBN design and spectral mask $\bDelta_{\thetad}$ must satisfy the constraints [C1] and [C2] over the entire beamtilt range $\thetad \in \cR_{\theta}$. 

\item[{[C4]}] The power amplifiers (PAs) must operate in a linear mode and their output power must be limited to a range $0 \, \mbox{dB} \leq |\vartheta_{k}(\thetad)|^{2} \leq 1 \, \mbox{dB}$.  

\item[{[C5]}] In order to minimize substrate loss and complexity in RFBN,  the number of stages inside RFBN is limited to 3. 

\item[{[C6]}] The incoming signals at the last stage of the RFBN combiner must be matched in amplitude and phase to account for insertion/microwave loss \cite{Pozar:microwave}.
\end{itemize}

The objectives are to (1) design the RFBN and DBF weights satisfying [C1 - C6] and (2) translate the designed RFBN weights into a microwave network minimizing microwave loss. We proceed with the RFBN design in the following order:
\bds
\item[{[P1a]}] We initially relax the microwave implementation constraints and PA efficiency. Given a specific architecture and performance requirements, how many transceivers do we really need?
	\bds
	\item[{[P1b]}] For an arbitrary $\Ntrx$ and limited dynamic range of PA, how can we design RFBN and DBF for optimal beampatterns? 
	\eds
\item[{[P2a]}] How do we represent the RFBN using microwave components such as power dividers and directional couplers (DCs)? What are the necessary conditions for optimal  RFBN factorizations?
\bds
\item[{[P2b]}] How does the design vary for macro-cellular and small-cell network?
\eds
\eds

The above two problems [P1] and [P2] form the core of this paper and their solutions are covered in the next three sections. Problem [P2] is subdivided, depending on the objectives of the cellular architecture, and a detailed synthesis and analysis of such architectures as well as RF design examples are provided in Sections \ref{sec:4} - \ref{sec:6}.

\section{Algorithms For Joint Optimization Of RFBN And DBF Weights}
\label{sec:3}
\noindent In this section, we consider problems [P1a] and [P1b], and estimate the RFBN and the DBF weights. 

\subsection{Bounds On The Number Of Transceivers}
\label{ssec:3_1}
The introduction of an RFBN reduces the order of the adaptive DBF to $\Ntrx$. Before we proceed to derive the RFBN and DBF weights, it is important to derive theoretical bounds on the minimum number of transceivers $\Ntrx$ satisfying [C1-C3] for a given $\cR_{\theta}$. We start with the MSE cost function in (\ref{eq:1_3}) and assume that we have obtained the optimal beamformer weights $\bu(\thetad)$ for the full dimension AAA. The design procedure for full dimension AAA is also specified in \cite{Palomar:joint_trx_beamforming, Yu:convex_beamforming}. For an ideal $\bu(\thetad)$ and cost function (\ref{eq:1_2}), we can approximate the result as 
\beq
\bDelta_{\thetad} 
\approx 
\bA(\theta)\bu(\thetad).
\label{eq:3_3a}
\enq

\begin{figure*}[t]
\beqa
\{ \bW, \, \bvartheta(\thetad) \} 
& \approx & 
\arg \min_{
\bW, \, \bvartheta(\thetad)
} \| \bA(\theta)\bu(\thetad) - \bA(\theta)\bW\bvartheta(\thetad) \|^2 
\quad 
\forall \, \thetad \in \{\cR_{\theta}\}  \quad \textrm{ since } \bDelta_{\thetad} 
\approx 
\bA(\theta)\bu(\thetad)
\nonumber \\
& \approx & \arg\min_{\{\bW, \, \bvartheta(\thetad)\}}
\|\bA(\theta)\|^{2} 
\| \, \bu(\thetad) - \bW \bvartheta(\thetad) \,  \|^2 
\quad 
\forall \, \thetad \in \{\cR_{\theta}\} 
\nonumber \\
& \approx & \arg\min_{\{\bW, \, \bvartheta(\thetad)\}}
\| \, \bu(\thetad) - \bW \bvartheta(\thetad) \,  \|^2 
\quad 
\forall \, \thetad \in \{\cR_{\theta}\} 
\label{eq:3_4}
\enqa
\end{figure*}

The joint DBF-RFBN optimization problem can be rewritten by plugging the LS approximation of $\bDelta_{\thetad}$ from (\ref{eq:3_3a}) in the original cost function (\ref{eq:1_3}). The modified optimization problem is written at the top of next page as (\ref{eq:3_4}).

The following Lemma characterizes the necessary conditions for optimal RFBN weights in the beamtilt range $\cR_{\theta}$.

\blm
Consider the scenario of [P1a]. Assume that the RFBN is made of ideal and lossless components and the PAs have infinite range. Given $\cR_{\theta} = \theta_{1}, \, \cdots, \, \theta_{N_{\theta}}$, the optimal weights of the RFBN must lie in the space spanned by the columns of dominant basis vectors of $\bTheta = [\, \bu(\theta_{1}), \, \cdots, \, \bu(\theta_{N_{\theta}})\,]$:
\[
\bW \in \mbox{col span} \{ \bTheta \}.
\]
\label{lemma:0}
\elm
\begin{proof}
Note that $\bW$ has to provide reasonable performance for all values of $\thetad \in \cR_{\theta}$. Stacking the cost function in (\ref{eq:3_4}) for the entire beamtilt range $\thetad \in \cR_{\theta}$:
\beqa
\{\bW, \bvartheta(\thetad)\} 
& = & 
\arg \min \|\left[ \bu(\theta_{1}), \, \cdots, \, \bu(\theta_{N_{\theta}})\, \right] \quad -  \nonumber \\
& & \quad \quad \quad \quad \quad \bW \left[ \bvartheta(\theta_{1}), \, \cdots, \, \bvartheta(\theta_{N_{\theta}})\, \right]\|^2 
\label{eq:3_5} \\
&\Leftrightarrow & 
\arg \min \|\bTheta - \bW\bVartheta\|^2 \nonumber
\enqa
where $\bTheta$ and $\bVartheta$ are respectively $\Nt \times N_{\theta}$ and $\Ntrx \times N_{\theta}$ matrices with $\bVartheta =  [\bvartheta(\theta_{1}), \, \cdots, \, \bvartheta(\theta_{N_{\theta}})\,]$. Let us assume that $\Nt \geq N_{\theta}$ and compute the singular value decomposition (SVD) of $\bTheta$:
\[
\bTheta = \left[\bU \right] \, \left[\bSigma\right] \, \left[\bV\right]^{H} = \left[ \bu_{1}, \, \cdots, \, \bu_{\Ntrx}, \,  \cdots \right]
\left[\bea{ccc}
\sigma_{1} & &\\
& \sigma_2 &\\
& & \ddots
\ena\right]
 \bV^{H},
\]
where $\bU$ and $\bV$ contain the left and right singular vectors and $\bSigma$ correspond to their singular values, typically arranged in descending order \cite{Moon:math_methods}. For $\Nt \geq N_{\theta}$, the optimal full dimension AAA weights $\bu(\thetad)$ can be approximated as a linear combination of the left singular vectors in $\bU$.  
If we have $\Ntrx$ transceivers, then choosing 
\[\bW = \left[ \bu_{1}, \, \cdots, \, \bu_{\Ntrx} \right]\] 
would provide the best $\Ntrx$-rank representation of $\bTheta$. 

For this reason, the RFBN design focussing on performance close to full dimension AAA is obtained by choosing $\Ntrx$ whenever $\sigma_{\Ntrx+1}$ tends to 0.
\end{proof}
A few remarks are in order:
\bds
\item From the array signal processing perspective, choosing dominant eigenvectors (usually in the digital domain) is referred to as reduced rank approaches \cite{vanVeen:beamforming},\cite{Goldstein:ms_wiener}. Such techniques are used to reduce digital post-processing complexity. 

\item Lemma \ref{lemma:0} assumes that $\Nt \geq N_{\theta}$. For $\Nt < N_{\theta}$, choosing $\Nt$ extreme values in $\cR_{\theta}$ and proceeding similarly will give an approximate $\bW$.

\item Lemma \ref{lemma:0} approach can be seen as a more systematic approach to estimate the weights of  Blass matrix, as in \cite{Mosca:Blass, Tarek:Nolen}. 

\eds
Note that this bound on the optimal $\bW$ does not consider microwave losses, linear operating range of PAs, and the number of possible interconnects in the overall network. 
\emph{However, it does provide a starting point for modifications in  next sections that consider the above practical issues.}

\subsection{RFBN optimization}
\label{ssec:3_2}

Estimating $\Ntrx$ for the required range of $\cR_{\theta}$ via Lemma \ref{lemma:0} is the first step in the RFBN design. Once we have established the minimum $\Ntrx$ for the desired SLL, the next step is the design of RFBN and DBF weights satisfying [C1-C6] and $\bDelta_{\thetad}$. 

We start from the cost function supplied in (\ref{eq:1_3}) and propose an interior-point algorithm \cite{Boyd:cnvx_opt} to jointly estimate $\bW$ and $\bvartheta(\thetad)$, which explicitly includes the constraints [C1]-[C3].

\subsubsection{RFBN optimization with constraints [C1-C3]}

One approach to design beamformers providing a main lobe at a specific direction while minimizing the overall variance along other directions is given by the the  Capon approach \cite{Moon:math_methods}.  Modifying the Capon approach for our DBF-RFBN setup, we can design the weights of $\bW$ such that the convolution of $\bW\bvartheta(\thetad)$ with the antenna array response $\bA(\theta)$ provides a mainlobe steered towards the desired sector, while minimizing the overall variance  of signal radiated from the array towards other sectors. 

Mathematically, the above two conditions can be combined and written using
$\bR_{\bA} = \bA^{H}(\theta)\bA(\theta)$ as
\beqa
\{\bW, \,\bvartheta(\thetad)\} & = &
 \min_{\{\bW\, \bvartheta(\thetad)\}} \,
\| \bvartheta^{H}(\thetad)\bW^{H} \bR_{\bA}\bW\bvartheta(\thetad) \|^2 
\label{eq:3_6} \\
& & \mbox{subject to} \quad \|\ba^{H}(\thetad)\bW\bvartheta(\thetad)\|^2=1.
\label{eq:3_6a}
\enqa

The coverage of signals from the antenna array towards the desired sector can be further enhanced by specifying the 3-dB or half power beamwidth ($\theta_{3, \, dB}$) constraint along the main lobe in the expression (\ref{eq:3_6}) i.e., 
\beq
\| \bvartheta^{H}(\thetad)\bW^{H} \ba(\thetad)\|^2 
= 1 
\quad 
\| \bvartheta^{H}(\thetad)\bW^{H} \ba(\theta_{3\, dB})\|^2 = 1/{2}.
\label{eq:3_6b}
\enq
Typically, $\thetad - \thetadB \leq 5^{\circ}$ in a macro-cell setup and $\thetad - \thetadB \leq 15^{\circ}$ in a small-cell setup. 

In order to suppress signals over unwanted sectors, we include SLL constraint [C2]. To achieve a specific SLL (say $\epsilon_{dB} = 20$ dB below the mainlobe) over a range of angles accounting for sidelobes $\bthetaSLL$, we introduce the constraint 
\beq
\| \bvartheta^{H}(\thetad)\bW^{H} \cA(\bthetaSLL)\|^2 \leq \epsilon
\label{eq:3_6sll}
\enq
where $\cA(\bthetaSLL)$ denotes the array response and $\epsilon = 10^{(-\epsilon_{dB}/10)}$ i.e. $\epsilon = 0.01$ for $\epsilon_{dB} = 20$ dB. 

\subsubsection*{Interior point optimization} Combining all the above constraints (\ref{eq:3_6a}) - (\ref{eq:3_6sll}), the central optimization problem becomes 
\beqa
\{\bW\, \bvartheta(\thetad) \}
& = &  \min 
\quad \bvartheta^{H}(\thetad)\bW^{H}\bR_{\bA}\bW\bvartheta(\thetad) \label{eq:3_9}\\
\forall \thetad \in \cR_{\theta} & & \mbox{subject to} \nonumber \\
& & \|\bvartheta^{H}(\thetad)\bW^{H}\ba(\thetad)\|^2 = 1 \,\quad \textrm{ from } (\ref{eq:3_6b})
 \nonumber \\
 & & \, \|\bvartheta^{H}(\thetad)\bW^{H}\ba(\thetadB)\|^2 = 1/2 \,\quad \textrm{ from } (\ref{eq:3_6b}) \nonumber\\
& & \|\bvartheta^{H}(\thetad)\bW^{H}\cA(\bthetaSLL)\|^2 
\leq [\epsilon, \cdots, \epsilon] \,\quad \textrm{ from } (\ref{eq:3_6sll}) \nonumber 
\enqa
where (\ref{eq:3_9}) specifies the main-beam as well as the SLL constraints. For the optimized $\Ntrx$ from Sec. \ref{ssec:3_1}, the above cost function explicitly includes the constraints [C1] - [C3]. The above cost function can be recast as a convex optimization problem \cite{Boyd:cnvx_opt} and solved numerically to obtain the optimal solution. The solution is obtained using the interior point algorithm \cite{Boyd:cnvx_opt}; note that similar techniques to estimate full dimension AAA weights $\bu(\thetad)$ have been proposed in \cite{Palomar:joint_trx_beamforming, Yu:convex_beamforming}.  
\newline
A few remarks are in order:
\bds
\item A more comprehensive approach is to jointly optimize $\bW$ and $\bvartheta(\thetad)$ , by representing $\bvartheta(\thetad)$ as a function of $\bW$ as in \cite{Vijay:apn_journal}
\[
\bvartheta(\thetad) \approx \bDelta_{\thetad} \, \left[ \bW \bA(\thetad) \right]^{\dagger}.
\]
\item  Note that the interior point optimization is not the main contribution of the paper and for a detailed performance analysis of these approaches, refer to \cite{Boyd:cnvx_opt, Palomar:joint_trx_beamforming}, \cite{Yu:convex_beamforming}.
\eds
 
\subsection{DBF Design for given RFBN satisfying [C4]}
\label{ssec:3_3}
Once  the optimal RFBN $\bW$ is designed as in \ref{ssec:3_2} for $\forall \thetad \in \cR_{\theta}$, the adaptive DBF weights are estimated for each beamtilt $\thetad$. Note that the DBF $\bvartheta(\thetad)$ is a function of the RFBN $\bW$ and  array response $\ba(\thetad)$. For a given $\bW$, designing DBF weights and minimizing the overall cost in (\ref{eq:3_4}) is transformed to 
\beqa
\bvartheta_{0}(\thetad) = \arg\min_{\bvartheta(\thetad)} \|\bDelta_{\thetad} - \bH(\theta)\bvartheta(\thetad)\|^{2}
\nonumber \\
\quad\mbox{where} \quad \bH(\theta) = \bA(\theta)\bW.
\label{eq:3_10}
\enqa

\subsubsection*{DBF design with PA constraints}
The PAs used for cellular networks are usually required by their ability to operate in a linear mode. Thus the gain and amplitude tapering with DBF should be limited to say 0 dB to 1 dB range. The DBF weights should comply with these output levels as specified in [C4]. We explicitly include these constraints on output power from each transceiver or DBF weights as 
\[
|\vartheta_{k}(\thetad)|^{2} \approx \frac{1}{\Ntrx} \quad \forall \, \, k \in \{1, \, \cdots, \Ntrx\}.
\]
For details on the optimization of DBF, refer to \cite{Yu:convex_beamforming}. Some comments are in order regarding the DBF design: 
\bds
\item The optimization can be expressed including the per PA power constraint as in \cite{Yu:convex_beamforming}. Note that for such algorithms to yield optimal solution, we need to explicitly show that the problem is convex. 
\item Please note that per antenna power constraint is not convex (unlike the inequality and linear constraints as proposed  in \cite{Yu:convex_beamforming}). As a special case, the expression $\vartheta_{k}(\thetad)$ can be represented using magnitude and phase terms and this magnitude constraint can be represented as a convex problem by exploiting the freedom to choose the phase. For further details on the applicability of such algorithms refer to \cite{Boyd:cnvx_opt}. 
\eds

\section{RFBN Implementation: Macro-cell network}
\label{sec:4}
Note that Section \ref{sec:3} provides some important directions on the design of RFBN, however, it does not represent the RFBN in terms of microwave components and account for practical limitations such as interconnect complexity and loss. The overall beamforming network, RFBN components as well as the design objectives vary for different scenarios and it is not possible to directly apply the results of Sec. \ref{sec:3} to design the RFBN. This section proposes design changes for specific architectures and factorizes the RFBN using a combination of microwave components.

\subsection{Two Stage Beamforming for cellular networks}
\label{ssec:4_1}

The DBF-RFBN arrangement can be seen as two-stage beamforming towards a specific sector. The first stage i.e. DBF is an adaptive transformation for each beamtilt with a straightforward implementation (say using FPGA as in our experimental setup). The second stage i.e. RFBN is made up of microwave components, and its implementation is not trivial, especially when the objectives are to minimize the overall loss and provide distinct beampatterns towards different sectors. 

The RFBN is comprised of a combination of commonly used microwave elements such as power dividers (Wilkinson dividers or WDs), phase shifters (micro-strip lines) and hybrid directional couplers (DCs) \cite[Ch. \, 7]{Pozar:microwave}. Typical implementations of coupler/divider elements result in loss of 0.1-0.2 dB. However most critical limitation in a dynamic RFBN setup is the insertion/return loss as we vary the DBF-RFBN setup for different beamtilts. The insertion loss occurs due to the amplitude and phase mismatch of the incoming signals at each combiner and depending on the beamtilt range of the RFBN,  and leads to decrease in power levels of up to  4.5 dB \cite{Mosca:Blass}. 

In addition, the RFBN has to be designed to account for a specific cellular arrangement. For example in a macro-cell  AAA setup, the range of beamtilt between adjacent sectors is small ($ \cR_{\theta} < 20^{\circ}$) and the distance between the mobile user and base station is typically large. For such scenarios, the emphasis is to design \emph{macro-RFBN} to achieve a narrow beam, focusing on minimizing the overall loss. Alternatively, in a small-cell  AAA setup, the beamtilt range is large ($60^{\circ} \, - \,90^{\circ}$) and the emphasis is on increasing the angular coverage of the AAA setup. In such a setup, the focus is more towards fixed 3-4 beams covering wide angular region. At this stage the joint design problem $\{ \bW, \, \bvartheta(\thetad)\}$ can be reclassified depending on the type of cellular architecture as 
\bds
\item[{[D1}] RFBN designed to minimize insertion loss over a relatively small beamtilt range, subsequently optimizing $\bvartheta(\thetad)$: This approach is typically suited for \emph{macro-cell} cases. 
\item[{[D2]}] RFBN designed to form arbitrary orthogonal beams, say at $\{+30^{\circ},\, 0^{\circ},\, -30^{\circ} \}$: This approach is typically suited for \emph{small-cell} cases.
\eds
Intutively [D1] and [D2] would lead to distinct redesigns of the RFBN algorithms proposed in Sec. \ref{sec:3}. We focus the rest of this section for the design of macro-cell $\bW$. The subsequent redesign of $\bvartheta(\thetad)$ is straightforward from Sec. \ref{sec:3} and hence omitted.

\bef
\begin{center}
	\includegraphics[width=\columnwidth]{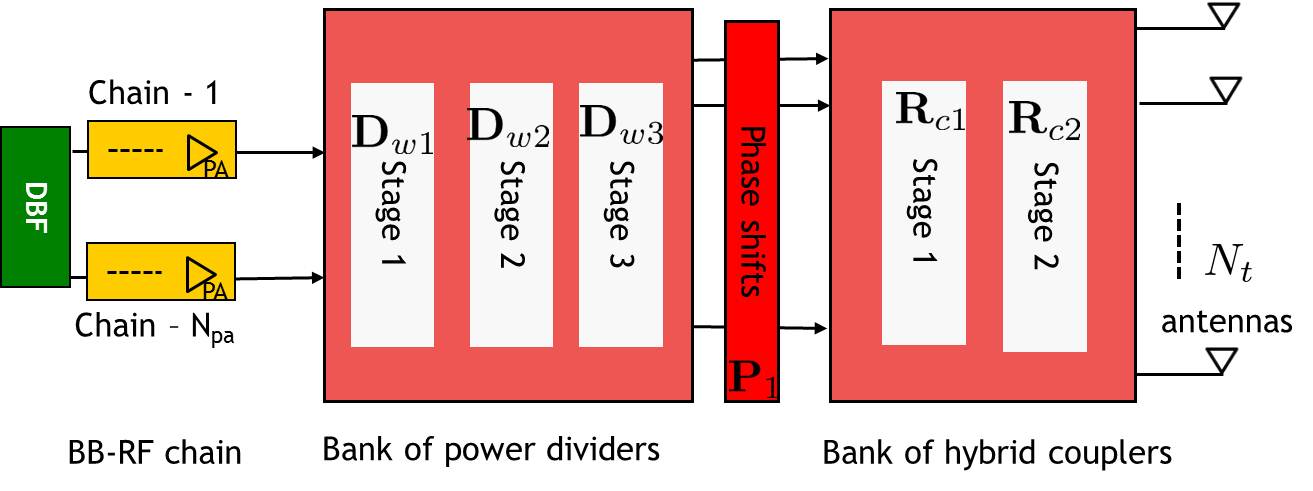}
\end{center}
\caption{\footnotesize{Factorization of an $\Nt \times \Ntrx$  RFBN into a bank of power dividers, phase-shifters, and DCs  connected to $\Nt$ antennas These stages are subsequently factorized into banks of 3-port or 4-port microwave networks. Specific decompositions depend on the cellular architecture (such as macro-cell or small-cell) and design objectives (such as insertion loss or beampattern).}}
\label{fig:2}
\enf

\subsection{Macro-Cell RFBN Synthesis}
\label{ssec:4_2}
We decompose the overall RFBN as shown in Fig. \ref{fig:2}, where components with similar functions are combined respectively into a filter-bank of power dividers $\bDfb$, a bank of phase shifters $\bP_{1}$ and a filter-bank of DCs $\bRfb$:
\[
\bea{ccc}
\bW & = & \bDfb \quad \times \quad \bP_{1}\quad  \times\quad \bRfb   \\
    & = & (\bD_{w\,1} \times \bD_{w\,2} \times \bD_{w\,3}) \quad\times\quad \bP_{1}\quad \times\quad (\bR_{c\,1} \times \bR_{c\,2}).
\ena
\]
In the above expression the subscripts, $\bD_{w\,i}$ denotes a filter-bank of Wilkinson dividers or power dividers for stage $i$ and $\bR_{c\,i}$ denotes a filter-bank of hybrid couplers or combiners for stage $i$. We will detail their functionalities in the subsequent sub-sections. 

\subsubsection{Redesign of $\bW$ to minimize insertion loss}
\noindent A generic $\Nt \times \Ntrx$ RFBN matrix  can be represented using a bank of 3-port networks containing a maximum of  
\bds
\item $(\Nt-1)$ power dividers connected to each transceiver i.e. $\Ntrx(\Nt-1)$ dividers in total.
\item $(\Ntrx-1)$ combiners connected to each antenna or $\Nt(\Ntrx-1)$ combiners in total.
\item $\Ntrx \Nt$ phase shifts to achieve the desired beampattern.
\eds
Such an arrangement would result in a significant increase in RFBN size and interconnect complexity. In addition, it is highly unlikely that the incoming signals at each DC will be matched in both amplitude as well as phase. This mismatch would increase the insertion loss further. 

\noindent \emph{Claim 1}: Consider scenario [P2], where the RFBN has been factorized into a bank of  DCs ($\bRfb$) as shown in Fig. \ref{fig:2}. Each bank is further divided into several stages of DCs $\bR_{c i}$. Irrespective of the RFBN setup and beamtilt range $\thetad$, an $\Ntrx\times 1$ vector $\bvartheta(\thetad)$ can be designed to minimize insertion loss at $\Ntrx-1$ combiners within $\bRfb$.

\noindent\begin{proof}
From linear estimation theory, we know that the $\Ntrx \times 1$ vector $\bvartheta(\thetad)$ has $\Ntrx$ degrees of freedom. From Lemma \ref{lemma:0} and (\ref{eq:3_5}), \emph{one} of the $\Ntrx$ variable in DBF can be used to achieve a mainlobe pointed towards $\thetad$. The amplitude and phase of the remaining $(\Ntrx-1)$ elements in DBF can be used to align the amplitude and phase at $(\Ntrx-1)$ combiners minimizing insertion loss. 
\end{proof}

Although Claim 1 specifies that one of the $\Ntrx$ degree is sufficient, the beampattern performance will be poor if we use only \emph{one} dimension among $\Ntrx$ dimensions to optimize for beampattern and the rest i.e. $(\Ntrx-1)$ dimensions are used to optimize for the insertion loss. Claim 1 acts as a starting point and provides a lower bound on the possible number of combiners for any beamtilt range minimizing loss in $\Ntrx-1$ ports. However, it makes sense to use all the available $\Ntrx$ degrees of freedom available in $\bvartheta(\thetad)$ to optimize for beampattern. This would mean a fundamental redesign of $\bW$ to minimize insertion loss. 

\subsubsection{Multistage RFBN decomposition}
\label{ssec:4_2_1}
The RFBN decomposition satisfying \emph{claim 1} and having at-most $\Ntrx-1$ combiners inside $\bRfb$ can be modeled using an $\Nt \times \Ntrx$ RFBN interconnect matrix $\bS$. One such example of the spatial interconnect map is, in a $11 \times 5$ case
\[
\bS = \underbrace{\left[\bea{ccccc}
\mathbf{1}_{3} & \mathbf{0}_{2} & \mathbf{0}_{4}& \mathbf{0}_{6} & \mathbf{0}_{4} \\
\mathbf{0}_{4} & \mathbf{1}_{3} & \mathbf{1}_{3}& \mathbf{1}_{3} & \mathbf{0}_{4} \\
\mathbf{0}_{4} & \mathbf{0}_{6} & \mathbf{0}_{4}& \mathbf{0}_{2} & \mathbf{1}_{3} \\
\ena\right]}_{11 \times 5 \textrm{  matrix}}
\]
where $\bones_l$ and $\bzeros_l$ respectively correspond to $l \times 1$ vector of ones and zeros. 
\newline \emph{Our focus is to redesign the RFBN weights satisfying $\bS$ as well as Lemma 1}. 

One approach to redesign RFBN is through the use of a modified version of orthogonal matching pursuit (OMP) \cite{Vijay:apn_journal} or through successive orthogonal projection of optimal $\bW$ \cite{Goldstein:ms_wiener}. We skip the exact algorithm details and briefly explain applying the algorithm for our RFBN setup as follows: 
\newline For a given $\bS$ as well as SVD of $\bTheta$:= $[\bU] \,[\bSigma]\, [\bV^H]$ 
\bds
\item From Lemma 1: $[\bu_{1}, \, \cdots \bu_{\Ntrx}, \, \cdots ] = \mbox{Basis}\{ \bTheta\}$.
\item for $ k \in \{1, \, \cdots, \Ntrx\}$
	\bds
			\item From right singular vectors of $\bTheta$: $ \bU= \left[ \bu_{1}, \, \bU_{N} \right]$.
			\item Extract  RFBN weights satisfying the spatial interconnects: $\bw_{k} = \bs_{k} \odot \bu_{1}$.
			\item Normalize each column of $\bw_{k}$. 
			\item $\bW = [\bw_{1}, \, \cdots \bw_{k}]$.
			\item Compute the orthogonal projection: $\cU = (\bI - \bW\bW^{H})\bTheta$.
			\item Update $\bTheta := \cU\bTheta$.
	\eds
\item end $k$.
\item Final RFBN: $\bW = [\bw_{1}, \, \cdots \bw_{\Ntrx}]$.
\eds
We perform RFBN decomposition  based on multi-stage Wiener decomposition \cite{Goldstein:ms_wiener}, since it provides a systematic low-complexity implementation of optimal $\bW$ and converges to optimal solution for increasing  $\Ntrx$. While the spatial interconnect map varies as we modify the constraints and claims, the methodology is generic. Note that we have skipped detailed synthesis and performance analysis of such approaches; for details refer to \cite{Goldstein:ms_wiener, Vijay:apn_journal}. 

\subsection{Macro-Cell RFBN analysis to minimize insertion loss}
\label{ssec:4_1a} 
\subsubsection{Factorization of power dividers $\bDfb$ and phase shifters $\bP_{1}$}
\label{ssec:4_2_2} 

\bef
\begin{center}
	\includegraphics[width=0.75\columnwidth]{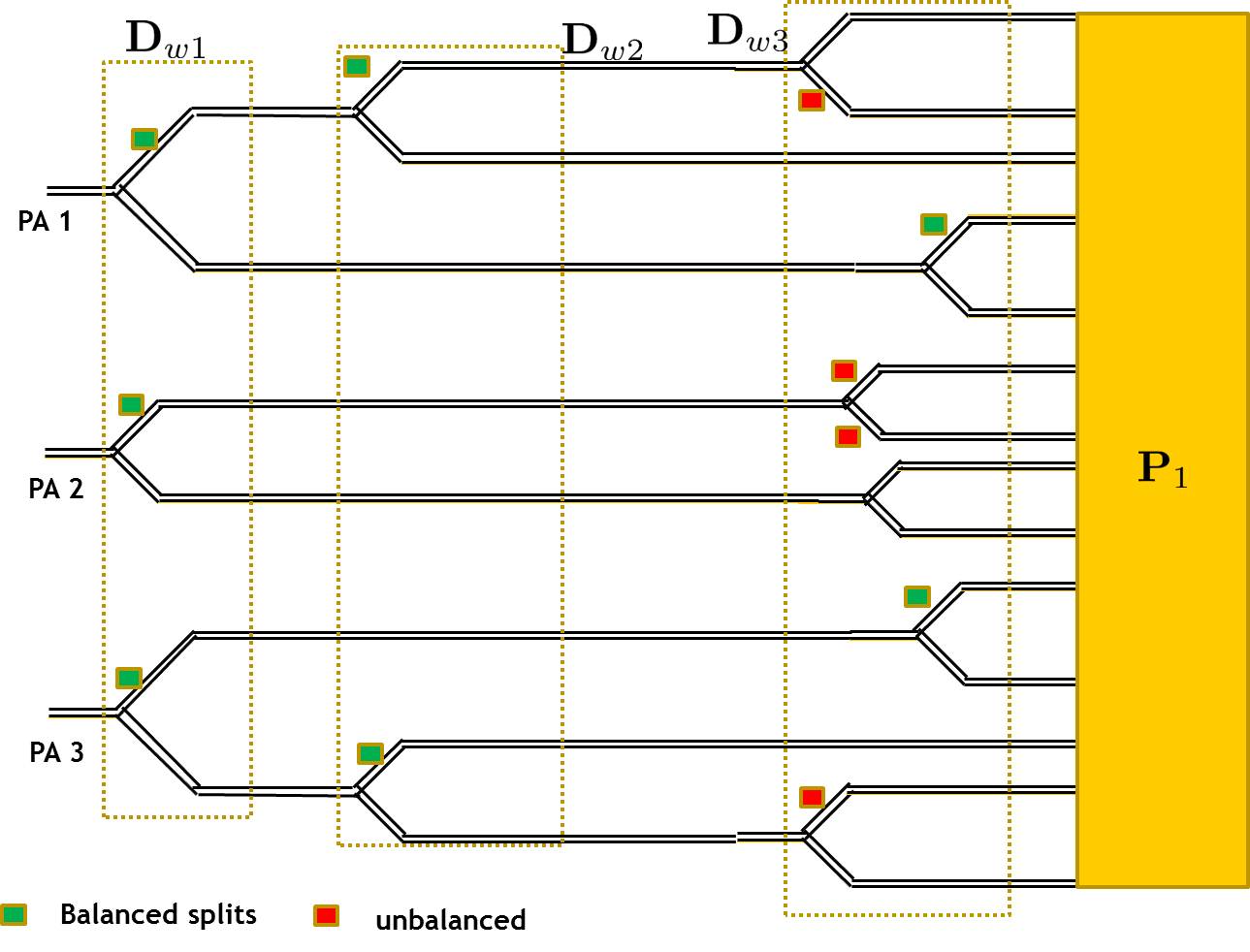}
\end{center}
\caption{\footnotesize{Power divider and phase shifter banks of macro-cell RF beamformer: Decomposition of power divider bank into balanced (green) and unbalanced Wilkinson power dividers (red). For a given $\Nt \times \Ntrx$ arrangement,  the unbalanced ratios of power dividers as well as the phase shifts are obtained by the multi-stage Wiener decomposition algorithms proposed in Sec. \ref{sec:3} and Sec. \ref{sec:4}.
}}
\label{fig:4}
\enf

\bef
\begin{center}
	\includegraphics[width=0.75\columnwidth]{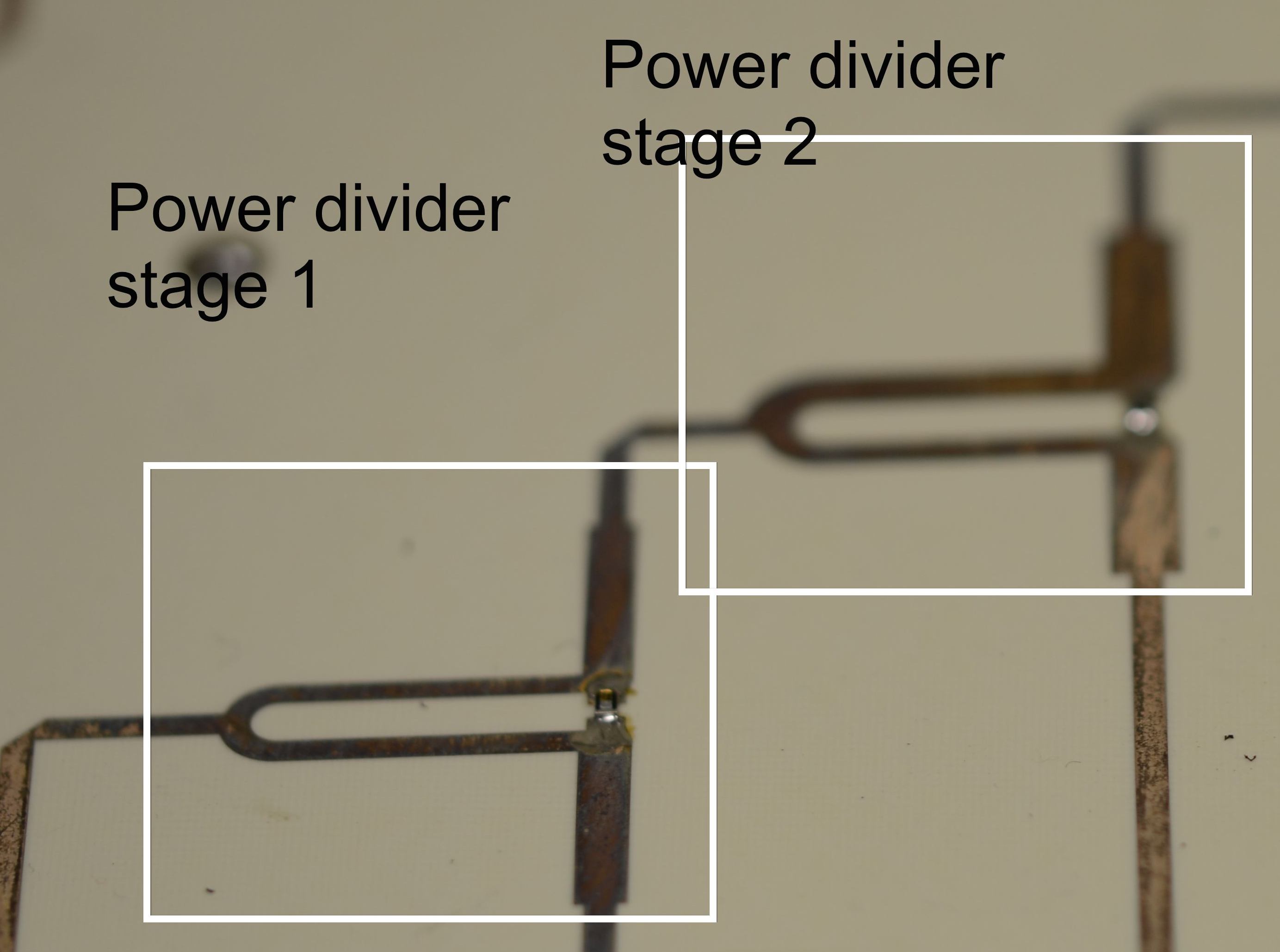}
\end{center}
\caption{\footnotesize{RFBN snapshot showing first two stages of power dividers. The power ratios and phase shifts are estimated using the multi-stage Wiener decomposition algorithm. The first stage contains balanced power dividers, and second stage unbalanced dividers. The output of second stage is fed to the phase shift matrix $\bP_1$.}}
\label{fig:4a}
\enf

The magnitude and phase values of column $i$ of $\bW$ i.e. $\bw_{i}$ correspond to power ratios and phase shifts of the output of the $i^\textrm{th}$ transceiver. Since $\Ntrx \ll \Nt$, the first stage of the RFBN consists primarily of power dividers to increase the number of input signals. In order to minimize the implementation complexity, each PA output is successively factorized into multiple stages of $\bDfb$ made up of 3-port WDs as shown in Fig. \ref{fig:4}. For reduced complexity, the design algorithms implement balanced WDs in the first two stages of $\bDfb$ followed by unbalanced WDs in the third stage. 
 
The output signals from $\bDfb$ undergo a phase shift denoted by diagonal matrix  $\bP_{1}$. The phase shifts are achieved by varying the lengths of the micro-strip lines. Note that the signals undergoing these phase shifts in $\bP_{1}$ correspond to that of $\bW$ and are already modified by the corresponding power ratios. At this point, the transmit signal $s[k]$ in (\ref{eq:1_1}) is modified by the DBF $\bvartheta(\thetad)$ and followed by $\bDfb$ and $\bP_{1}$ to obtain the achieve the desired beamtilt and pattern.

Fig. \ref{fig:4a} shows a segment of RFBN with two stages of power dividers followed by  phase shifters. The power ratios and and phase shifts are obtained from Sec. \ref{ssec:4_2_1}. Fig. \ref{fig:4b} shows an unbalanced divider, whose power ratios given by $\bw_i$ are proportional to the width of the power divider arms.  Similarly, the strip-line lengths at the power divider output are unequal, and  correspond to the phase shift specified in $\bw_i$.Let $N_{\bs_{i}}$ be the number of non-zero elements in the column of $\bw_{i}$. At the output of phase shifter, the total number of signals are $N_{\bs} = \sum_{i=1}^{\Ntrx}N_{\bs_{i}}$. Note that it is necessary to have $N_{\bs} \gg \Nt$ to achieve the desired beampattern for various beamtilts. 

\bef
\begin{center}
	\includegraphics[width=0.5\columnwidth]{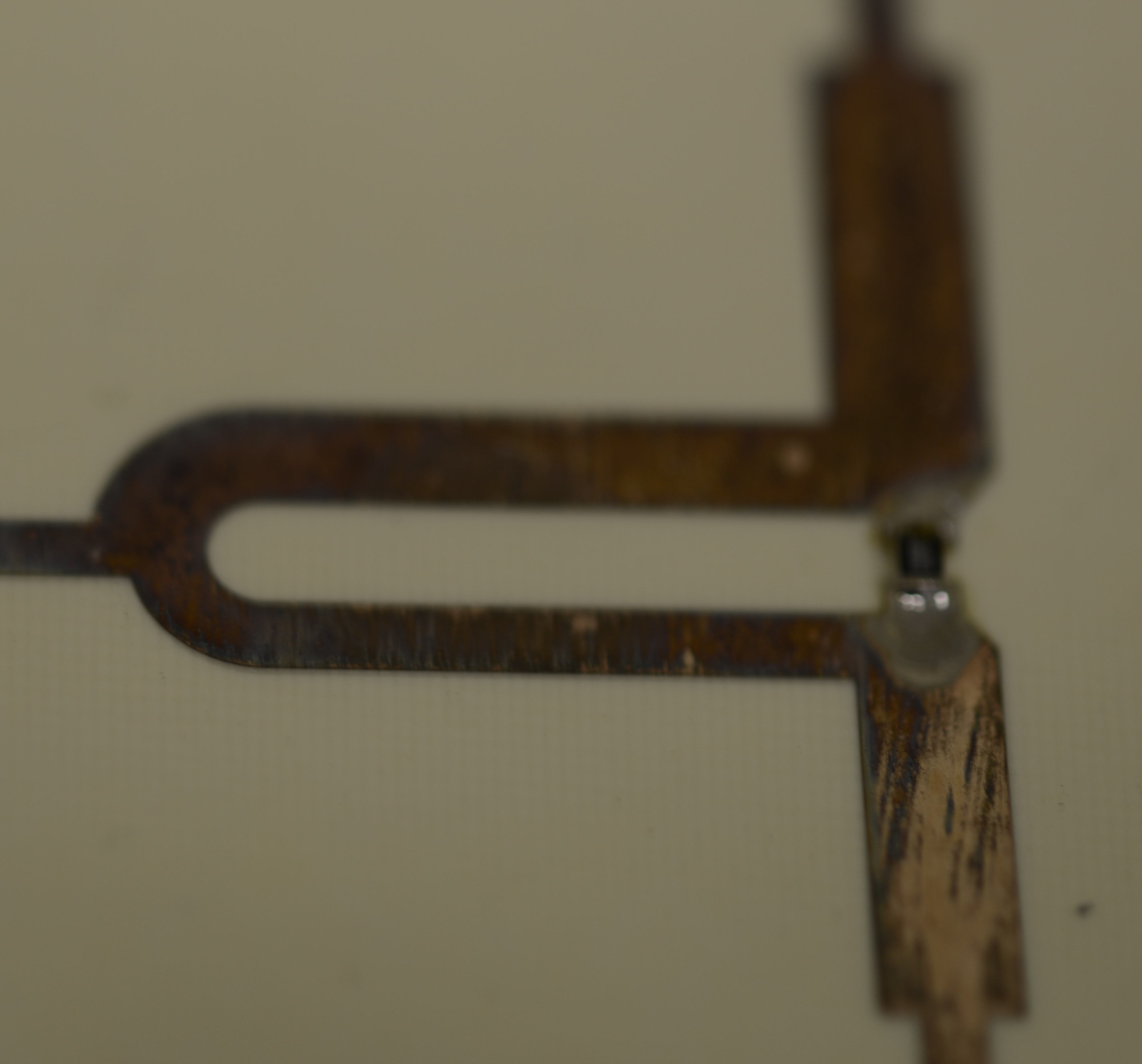}
\end{center}
\caption{\footnotesize{RFBN snapshot detailing unbalanced power divider ratios.}}
\label{fig:4b}
\enf

\subsubsection{RFBN as a linear phase finite impulse response (FIR) filter}
The $N_{\bs}$ outputs from $\bP_{1}$ are modified and combined by $\bRfb$ and fed to $\Nt$ antennas. For efficient and lossless operation of $\bRfb$, it is necessary that the input signal at each power combiner/DC be matched in terms of amplitude and phase and any mismatch will result in insertion loss. Note that the RFBN is fixed, but the phase and amplitude of $\bvartheta(\thetad)$ is varied for each beamtilt. Such a dynamic arrangement when used with the standard 3-port DCs will always result in insertion loss. 

For this reason, we use hybrid elements, such as rat-race couplers or branch hybrids (four port networks) \cite[Ch. \, 7]{Pozar:microwave} instead of a standard three port combiner. In a rat-race coupler,  Ports 2 and 3 are input ports and the inputs are coupled to a standard output (Port 1). A fourth port (also referred to as the reflection port or isolation port) extracts the signals that would have otherwise led to insertion loss whenever there is a mismatch between the input signals \cite[Pg. 480]{Pozar:microwave}. Thus, at a given stage $R_{c\, i}$, any phase or amplitude mismatch can be captured at the output of the isolation port of the hybrid coupler. We exploit the following signal processing properties of such hybrid couplers  to reroute the insertion loss seen in Stage $i$ and minimize the overall loss in subsequent stage $i+1$. 
\bef
\begin{center}
	\includegraphics[width=0.9\columnwidth]{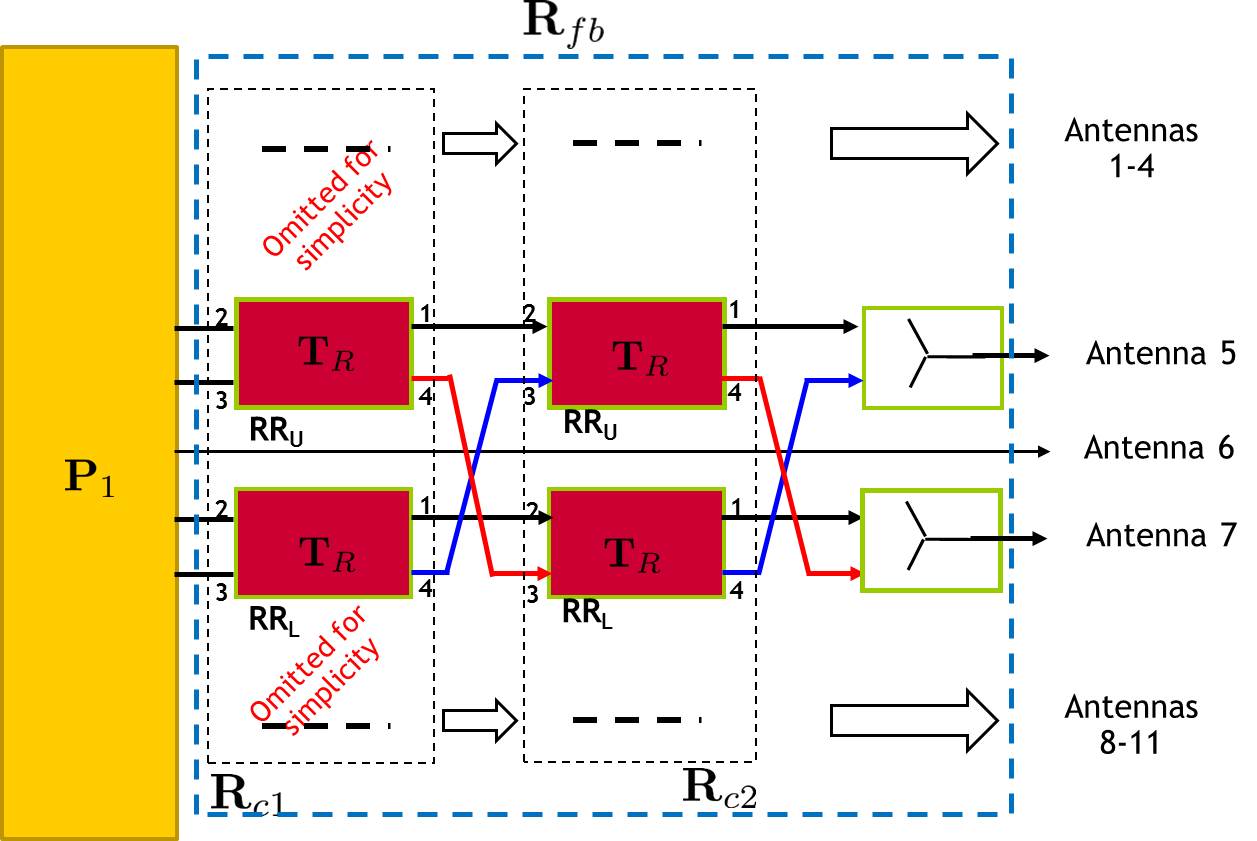}
\end{center}
\caption{\footnotesize{Directional coupler bank $\bRfb$ for macro-cell RF beamformer factorization: For $11 \times 5$ instance, each stage is made-up of a two 4-port rat-race couplers (denoted by $2 \times 2$ matrix $\bT_{\textrm{R}}$). These couplers are used to combine RF signals from adjacent ports. The combined signals are subsequently recirculated at the next stage to compensate for insertion losses. For simplicity, only the combiners connecting Antennas 5 and 7 for a $11 \times 5$ setup is detailed.}}
\label{fig:5}
\enf
\bef
\begin{center}
	\includegraphics[width=0.5\columnwidth]{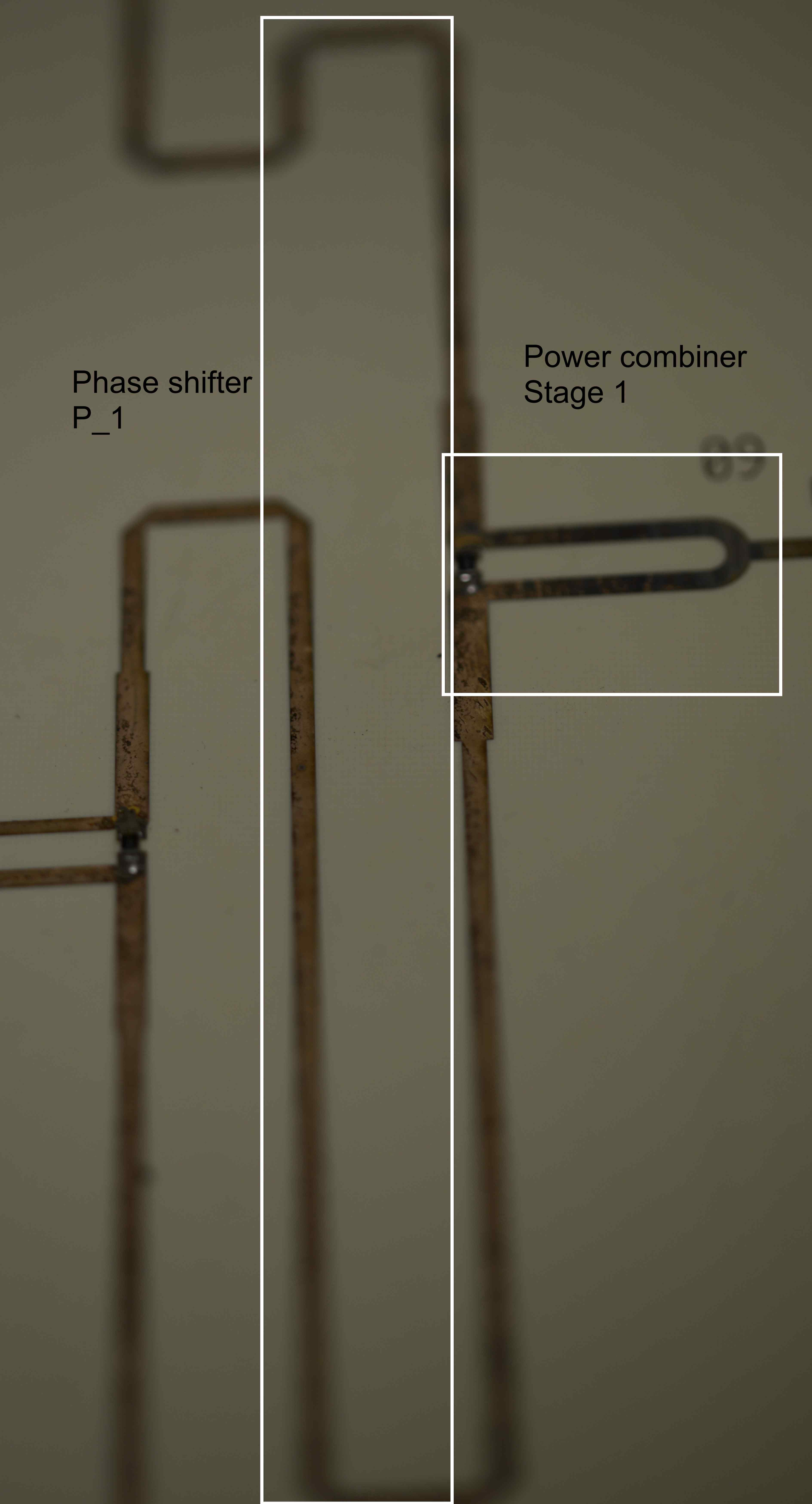}
\end{center}
\caption{\footnotesize{RFBN snapshot with phase shifter matrix $\bP_1$ implemented by micro-strip lines followed by first stage of $\bRfb$.}}
\label{fig:5a}
\enf

\emph{Claim 2}: Consider scenario [P2], where the RFBN is subdivided into 4-port DCs. 
\bds
\item The elements of the combined response of $\bW\bvartheta(\thetad)$ providing the main lobe along $\thetad$ match the impulse response of full dimension AAA.
\item The insertion loss in each stage of $\bR_{c \, i}$ is minimized, if the number of hybrid couplers does not exceed $\Ntrx-1$.
\eds

\begin{proof}
For simplicity, consider the full dimension AAA setup with $\Nt$ transceivers and array response $\bA(\theta)$ as specified in (\ref{eq:1_2}) and the full dimension AAA beamformer $\bu(\thetad) = [u_0(\thetad), \, \cdots, \, u_{\Nt-1}(\thetad)]^{T} $. Note that $\ba(\theta_{i})$ has linear phase progression since the antenna elements are uniformly spaced. The transfer function of $\bu(\thetad)$ operating on the antenna array can be written as 
\beq
U(\theta_{i}) 
= 
\sum_{k=0}^{\Nt-1}u_{k}(\thetad) \, e^{-j2\frac{\pi}{\lambda}k \cos(\theta_{i})} 
\quad \forall \theta_{i} \in \,\, [-\pi, \pi].
\label{eq:4_12}
\enq
The full dimension beamformer $\bu(\thetad)$ can be seen as a spatial FIR filter operating on the antenna array and the above expression (\ref{eq:4_12}) can be seen as the spatial equivalent of a filter transfer function or filter spectral response. 

From filter design theory, we know that matched filters operating on uniformly sampled antennas will have a \emph{symmetric} magnitude response along the central antenna element $\Nt/2$. They will also have a linear phase progression: 
\[
\bea{ccc}
|u_{k}(\thetad)| & =& |u_{\Nt-k}(\thetad)| \quad \textrm{and} \\
\quad \angle{u_{k}(\thetad)} - \angle{u_{k-1}(\thetad)} & =& \angle{u_{k+1}(\thetad)} - \angle{u_k(\thetad)}.
\ena
\]

We extend this \emph{linear phase} and \emph{symmetric magnitude} argument for our RFBN-DBF setup $\bW\bvartheta(\thetad)$ operating on $\bA(\theta)$ and its  transfer function is similar to $U(\theta_{i})$  in (\ref{eq:4_12}).

 The symmetric linear phase property ensures that the elements of $\bW\bvartheta(\thetad)$ as well as the columns of $\bA(\theta)$, will have linear phase progression. Exploiting the shift invariance property, the combined DBF-RFBN $\bW\bvartheta(\thetad)$ operating on antenna array $\bA(\theta)$ will also have a linear phase progression. This linear phase progression throughout the network will ensure the isolation port signals at stage $i$ of $R_{c,\,i}$ will be in phase with the input (port 1 signal) at stage $R_{c,\,i+1}$.

Due to linear phase progression, the signal from the isolation port of the rat-race Coupler $k$ is in phase with respect to the signal at the output port of the rat-race Coupler $\Ntrx-k \, \, \forall , k \in \{1, \, \cdots, \Ntrx-1\}$. This property means that the above two signals from \emph{stage} $i$ can be combined to mitigate the insertion loss at the subsequent stage $\bR_{c, \, i+1}$ as shown in Fig. \ref{fig:5}. 
\end{proof}
Thus, the combined matched filter $\bW \bvartheta(\thetad)$ will maximize the SNR for given beamtilt, provide optimal beampattern and minimize insertion loss. 

\subsubsection{Factorization of hybrid couplers $\bRfb$}
\label{ssec:4_2_3}

The hybrid elements can either be branch hybrids (commonly used in Butler matrix implementations \cite{Butler:butler_matrix}) or rat-race hybrids. In our setup, we use rat-race hybrid elements, the motivation being
\bds
\item Four-port rat-race couplers can also be seen as radix-2 discrete Fourier transform (DFT) implementations. Higher order DFTs can then be obtained using a bank of such couplers.
\item The DFT analysis provides a generic approach to construct $\bRfb$  using a careful arrangement of rat-race couplers at different positions to mitigate the overall loss. For example, the arrangement of rat-race couplers in $\bRfb$ as in Fig. \ref{fig:5} provides a linear phase progression and symmetric magnitude response. 
\eds

\section{RFBN Implementation: small-cell Network}
\label{sec:4b}
In a small-cell setup, the number of RF chains are limited to 2 or 3 and the number of antennas are limited to 4-6 (typically as an horizontal arrangement). The objective is to provide coverage along $N_{\theta} = 3 - 4$ fixed beams spaced $30^{\circ}$ apart from each other (say $\thetad \in \{-30^{\circ}, \, 0^{\circ}, \, +30^{\circ}\}$). Each beam has a wider 3-dB beamwidth (nearly $15^{\circ}$ and the  focus is more on improving angular coverage (unlike the macro-cell case where the focus is on minimizing loss). The requirements and overall setup make this design fundamentally different from that of Sec. \ref{sec:4}. 

\subsection{Connections with state of the art}
\label{ssec:4_3_1}
Existing passive beamformers such as \cite{Butler:butler_matrix}  provide distinct (and orthogonal) beams while minimizing microwave loss. The setup \cite{Butler:butler_matrix} can be modeled having $\Nt$ inputs and $\Nt$ outputs, connecting $\Nt$ transceivers and $\Nt$ antennas and implemented using either branch or rat-race hybrid couplers. Typical implementations have $\Nt=4$ antennas spaced half a wavelength ($\lambda/2$) apart. Note that in the majority of such lossless implementations, only one transceiver is turned \emph{on} at a given time to generate the desired beampatterns. Such a design sacrifices radiated power from the array for lossless implementation.

A systematic approach to estimate the RFBN weights for arbitrary beamtilts $\cR_{\theta} = \{\theta_{1} , \theta_{2},\, \theta_{3} \}$  can be obtained from the basis vectors of $\bTheta$ using a QR decomposition or SVD as specified by Lemma 1. The QR decomposition approach is similar to the Blass Matrix design of \cite{Mosca:Blass}. In \cite{Mosca:Blass}, the authors propose an RF matrix whose columns correspond to orthonormal vectors of $\bTheta$. However they do not account for operational constraints [C1]-[C3]. subsequently, in order to minimize the insertion loss, they transform the Blass matrix into a modified Butler matrix, with only one PA operating at a time. This modification sacrifices the effective radiated power from $\Ntrx$ transceivers. Improving from the the Blass Matrix design of \cite{Mosca:Blass}, Djerafi, \emph{et. al.} \cite{Tarek:Nolen} propose an RF beamforming matrix that provides a set of orthogonal beams with multiple PAs operating at the same time. Thus the existing family of RFBNs to generate distinct/orthogonal beams can be classified into 
\bds
\item Lossless Butler matrix implementations \cite{Butler:butler_matrix}, where there is only one PA operating at a  time, sacrificing on the overall radiated power. 
\item Lossy Blass matrix \cite{Mosca:Blass, Tarek:Nolen, Peng:Blass} with $\Ntrx$ transceivers operating simultaneously, sacrificing insertion loss performance for effective radiated power.
\eds
The fundamental question is whether we can achieve a hybrid combination of existing designs, generating orthogonal beams and satisfying [C1-C6], while keeping insertion loss to manageable levels.  

\subsection{Generalized Butler matrix }
\label{ssec:4_3_3}

If we implement the RFBN based on \cite{Mosca:Blass, Tarek:Nolen},  $\bRfb$ will have more than $\Ntrx-1$ combiners and \emph{claims} 1-2 will not be satisfied.  Additionally, the DBF-RFBN arrangement will not always have a linear phase progression due to limited degrees of freedom.

Our objective is RFBN design satisfying orthogonal beamtilts $\cR_{\theta}$ while keeping the insertion loss to manageable levels. We start from the optimal RFBN design in Sec. \ref{ssec:3_2}. 

Given an $\Nt \times \Ntrx$ RFBN, the signal at the antenna element $i, \, i \in \{1, \, \cdots, \, \Nt\}$ can be routed through $\Ntrx$ combiners. These correspond to the number of non-zero elements of the the row $\bW(i, :)$ and we denote this quantity by \emph{row-weight}. One approach to reduce insertion loss is to match the combining signals in amplitude and phase. In cases where this matching is not possible, it is preferable reduce row-weight without modifying the overall DBF-RFBN response. 

From the RFBN designed in  Sec. \ref{ssec:3_2}, we search for the unmatched/out-of-phase combiners:
\bds
\item Selectively remove the connections corresponding to these unmatched combiners:  This operation can be done by zeroing specific entries in the RFBN matrix that exceed a specific mismatch threshold. To zero a specific entry in the RFBN matrix, we use the \emph{Givens Rotation} method \cite{Moon:math_methods}.
\item Subsequently, the updated RFBN is used to re-estimate DBF  $\bvartheta(\thetad)$ as in Sec. \ref{sec:3}-B such that the input signals at the remaining combiners are matched and constraints [C1]-[C6] are satisfied.  
\eds

As an example, we illustrate the Givens rotation based RFBN design for $\bvartheta(\theta_{1}) = [1, \, 1,  \, 1]^{T}$, say at beamtilt $\theta_{1}$. The combined DBF-RFBN response can be represented using arbitrary complex variables $\ast$ as  
\beq
\bW \,  \bvartheta(\theta_{1}) = 
\underbrace{\left[
\bea{c} 
\left[\bea{ccc}
\ast & * & *	\\
\ast & * & *	\ena\right]\\
\ast \leftarrow \fbox{*} \rightarrow *		\\
\left[\bea{ccc}
\ast & * & *	\ena\right]
\ena\right]}_{\bW}
\underbrace{ \left[\bea{c}1 \\ 1 \\ 1 \ena\right]}_{\bvartheta(\thetad)}
.
\label{eq:4_2}
\enq
Consider above expression, where $w_{3, \, 2}$ (element at row 3 and column 2) is out of phase with the two other signals $w_{3, \, 1}$ and $w_{3, \, 3}$. Givens rotation allows us to remove a specific element (in this case $w_{3, \, 2}$) using the  $\Nt \times \Nt$ matrix for estimated phase $\phi$
\[
\bG(2, 3, \phi) = \left[\bea{cccc}
1 & & & 	\\
& \cos(\phi) & \sin(\phi) &\\
& -\sin(\phi) & -\cos(\phi) &\\
& & & 1
\ena\right]  
\]
where $\phi\, \in \{-\pi, \, \pi\}$. 

A generalization of the above expression $\bG(i, j, \phi)$ is recursively applied to zero elements along the $i^{\textrm{th}}$ row and the $j^{\textrm{th}}$ column, whenever the amplitude and phase mismatch exceeds a particular threshold. While doing this operation, we need to make sure that the overall transfer function of the matrix does not change considerably. Thus we have
\beqa
\bW
& :\Rightarrow& 
\bG(3,2,\phi_{1})\bW 
= 
\underbrace{\left[\bea{ccc}
\ast & * & \fbox{*}	\\
\ast & * & *	\\
\ast & 0 & *	\\
\ast & * & *	
\ena\right]}
_{\textrm{recursion 1} } :\Rightarrow \nonumber \\
& :\Rightarrow &
\bG(3,1,\phi_{2})\bW =
\underbrace
{
\left[\bea{ccc}
\ast & * & 0	\\
\ast & * & *	\\
\ast & 0 & *	\\
\ast & * & *	
\ena\right]}_{\textrm{recursion 2}}
\enqa
In practice, we perform successive Givens rotation on the \emph{combined response} $\bW\bVartheta$, where $\bVartheta = [\bvartheta(\theta_{1}), \, \bvartheta(\theta_{2}), \, \bvartheta(\theta_{3})]$. Some comments are in order:
\bds
\item Note that after every rotation, the DBF $\bvartheta(\thetad)$ as well as the combined response $\bW\bvartheta(\thetad)$ has to be re-estimated to account for SLL requirements, PA limitations, etc. 
\item The number of Givens rotation iterations is a tradeoff between the quality of beampatterns (desired SLL/beamtilts) and the insertion loss levels.For example, increasing the number of iterations might degrade the beam-pattern performance.
\item Bulter matrix can be interpreted as a special case of Givens rotation, where $\bVartheta$ is an $4 \times 4$ identity matrix whenever $\Nt = \Ntrx=4$.
\eds

\section{Simulation results}
\label{sec:5}
To assess the performance of the joint RFBN-DBF architectures, we have applied it to macro and small-cell multi-antenna base stations. We present simulation results for the RFBN setup based on multi-stage Wiener decomposition as outlined in Sec. \ref{sec:4}. These results include computing the beampatterns for varying number of transceivers, varying beamtilts and estimating the sidelobe level and insertion loss performance. These results are complimented by RFBN implementation and measurements detailed in Sec.\ref{sec:6} and Sec. \ref{sec:7} respectively. The performance indicators are usually:
\bds
\item The radiated energy along desired beamtilt $\thetad$ and sidelobe levels.
\item insertion loss and beamtilt range limitations with RFBNs.
\eds
The joint RFBN-DBF arrangement should be able to beamform and transmit the desired signal towards distinct sectors $\thetad \in \{0^{\circ}, \, \cdots, 20^{\circ}\}$. We consider a setup with $\Nt=10-12$ antennas radiating at 2.6 GHz. The objective is to ensure SLLs are 18-20 dB below main-lobe. The antenna element has a 3-dB beamwidth $65^\circ$ and complies with the 3GPP specifications \cite{3GPP:LTE}.  Antenna elements are uniformly spaced at a distance $0.8 \lambda$. Note that Nyquist sampling criterion would typically limit array spacing to $0.5 \lambda$ at the given frequency of operation, however cellular antennas are typically designed for wide-band operation and the spacing constraint is relaxed. The increased spacing leads to grating lobes for some beamtilts. 

Lemma \ref{lemma:0} says that we need at least $\Ntrx=3$ transceivers for a macro-cell scenario with beamtilt range to achieve 18-20 db SLL for the entire beamtilt range $\thetad \in \{0^{\circ},\,\cdots,\,20^{\circ}\}$. In practice, we would need 4-5 transceivers to additionally account for insertion loss, limited dynamic range of the PAs and achieve desired array gain. 

\bef
\begin{center}
\includegraphics[width=\columnwidth]{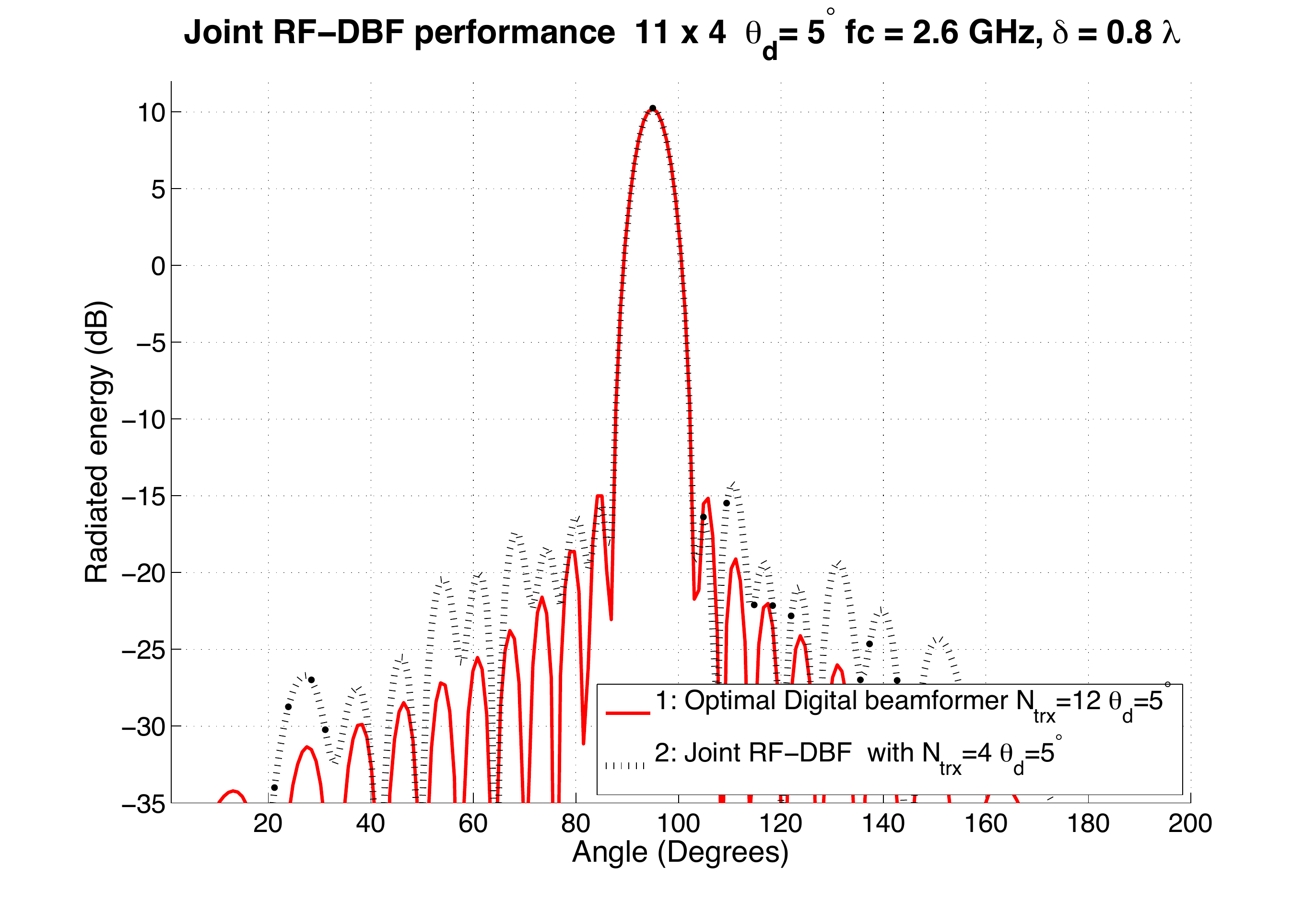}
\includegraphics[width=\columnwidth]{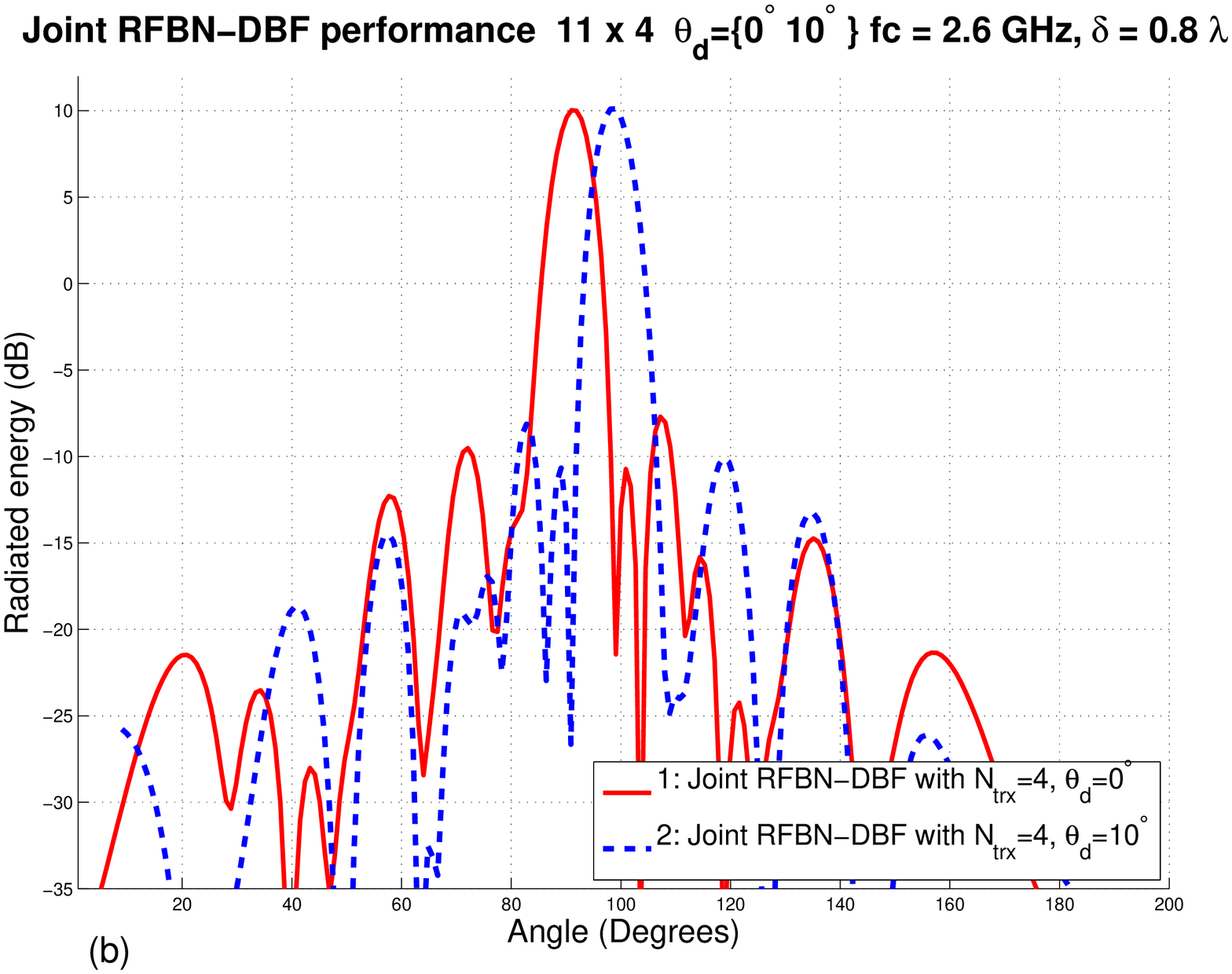}
\end{center}
\caption{\footnotesize{Joint RF- and digital beamformer (DBF) performance with  $\Nt=11$ antennas for multiple beamtilts $\thetad$: (a) Comparison of RFBN-DBF performance with optimal DBF performance when used with a full dimension AAA setup containing $\Nt=\Ntrx=11$ transceivers (b) Performance of joint RFBN-DBF with non-adaptive RF beamformer plus varying Digital beamformer for $\thetad=\{0^{\circ},\,10^{\circ}\}$}}.
\label{fig:7}
\enf
\bef
\begin{center}
\includegraphics[width=\columnwidth]{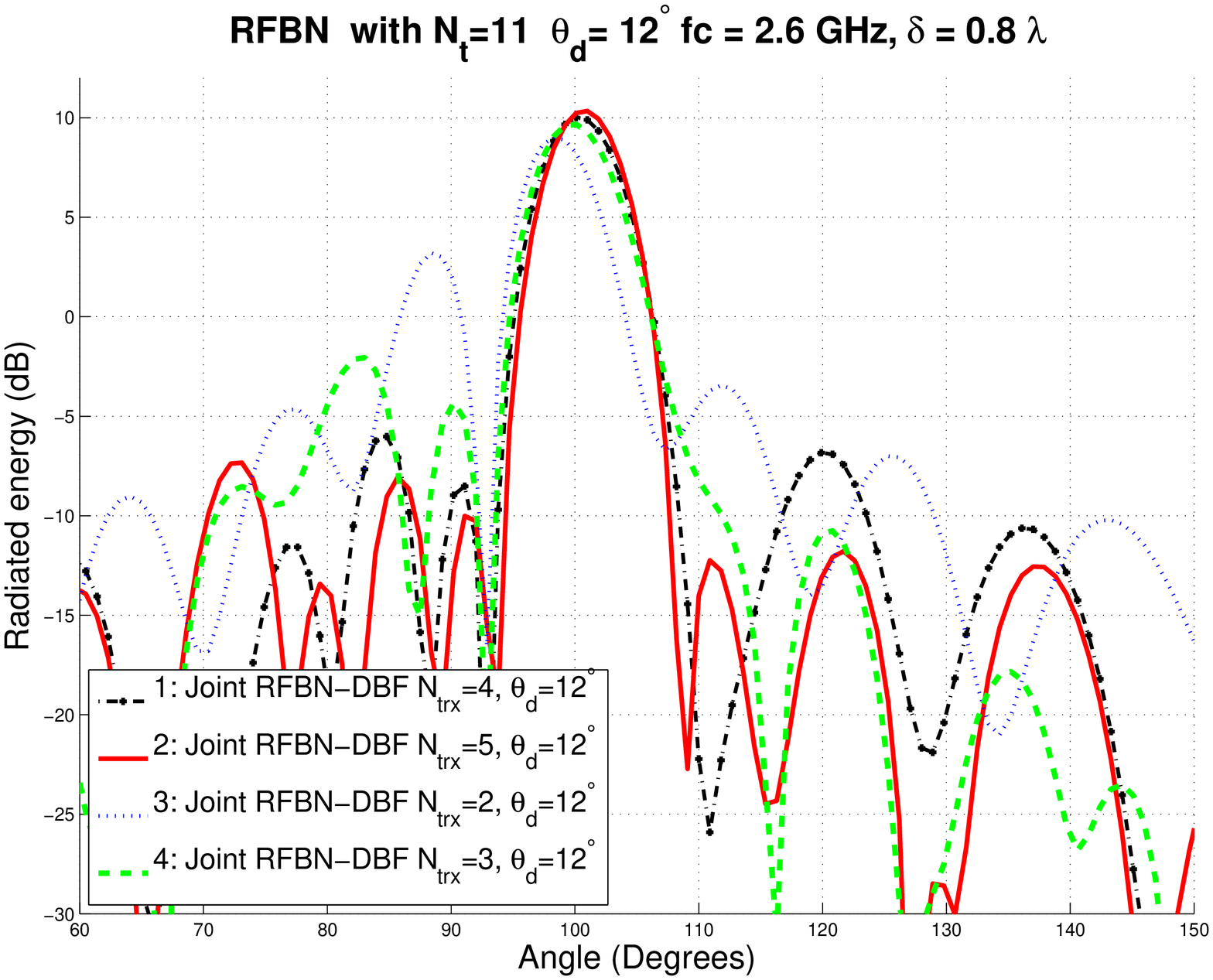}
\end{center}
\caption{\footnotesize{Optimal Joint RFBN-DBF performance for different transceiver configurations for fixed beamtilt $\thetad=12$ and $\Ntrx=2,\, 3,\, 4 \mbox{ and } 5$. Notice that the side-lobe levels degrade as $\Ntrx$ reduces}}
\label{fig:8}
\enf

\subsection*{Vertical sectorization performance}
\label{ssec:5_1}

Fig. \ref{fig:7}(a) shows the beampattern performance for an $\Ntrx=4$ transceiver and $\Nt=11$ antenna setup. Curve 1 is the reference curve corresponding to optimal beamformer with $\Ntrx=11$ transceivers with $\Ntrx=11$ PAs operating with infinite dynamic range. Note that this is unrealistic in practice and is shown exclusively as the theoretical bound for optimal beamformer design. Curve 2 shows the performance with an $11 \times 4$ RFBN and an $4\times 1$ DBF. The partially adaptive setup provides optimal main-lobe performance while satisfying the 3GPP requirements and  converges to curve 1. Both approaches provide 24 dB SLL and $\theta_{3, \, \mbox{dB}} = 5^{\circ}$ for a radiated power of 10 dB along  $\thetad = 5^{\circ}$.

Fig. \ref{fig:7}(b) shows the performance for two sectors  $\thetad = 0^{\circ} \, \mbox{and} \, \thetad=10^{\circ}$. For clarity, we have shown only two beamtilts, but the setup can account for the entire range $\thetad \in \{0^{\circ}, \, \cdots, \, 20^{\circ} \}$. Comparing the curves 1 and 2, we observe that the \emph{partially adaptive} RFBN-DBF achieves 18.5 dB SLL performance. Note that the radiated power along $\thetad$ is still preserved. 

Fig. \ref{fig:8} shows the beampatterns for varying RFBN arrangements with $\Ntrx=\{2, \, \cdots, \, 5\}$ and $\Nt=11$. In this case, $\Ntrx$ is fixed in each case and subsequently the optimal RFBN is designed.  The RFBN is initially optimized for the sectors $\cR_{\theta} \in \{0^{\circ}, \, \cdots, 10^{\circ}\}$. Curves 1-4 show a snapshot of beampattern performance, when we are required to provide a main lobe across $\thetad=12^{\circ}$ outside $\cR_{\theta}$. Curves 1 ($\Ntrx = 4$) and 2 ($\Ntrx = 5$) show reasonably good performance with 16 and 18 dB SLL respectively, while achieving $\thetadB=5^{\circ}$. As expected, the performance significantly degrades for an RFBN arrangement with $\Ntrx=2$. To account for design flexibility and address different $\thetad$, it is necessary to keep $\Ntrx \geq 3$.

\section{Network Design Examples}
\label{sec:6}
The two-stage DBF-RFBN approach must be eventually be realised using a microwave circuit. This section detains the microwave design of one such architecture and analyse its performance. The DBF implementation is straightforward and is omitted for simplicity. We use Advanced Design Systems (ADS) from Agilent Technologies \cite{Agilent:ads} to implement the RFBN and observe its performance. As mentioned in Sec. \ref{sec:4}, the main difference between a macro-cell application and small-cell application is that the former needs to provide a continuous tilt of the main beam within a certain range, whereas the latter provides orthogonal set of beams.
\begin{figure*}
\begin{center}
\includegraphics[width=0.8\textwidth]{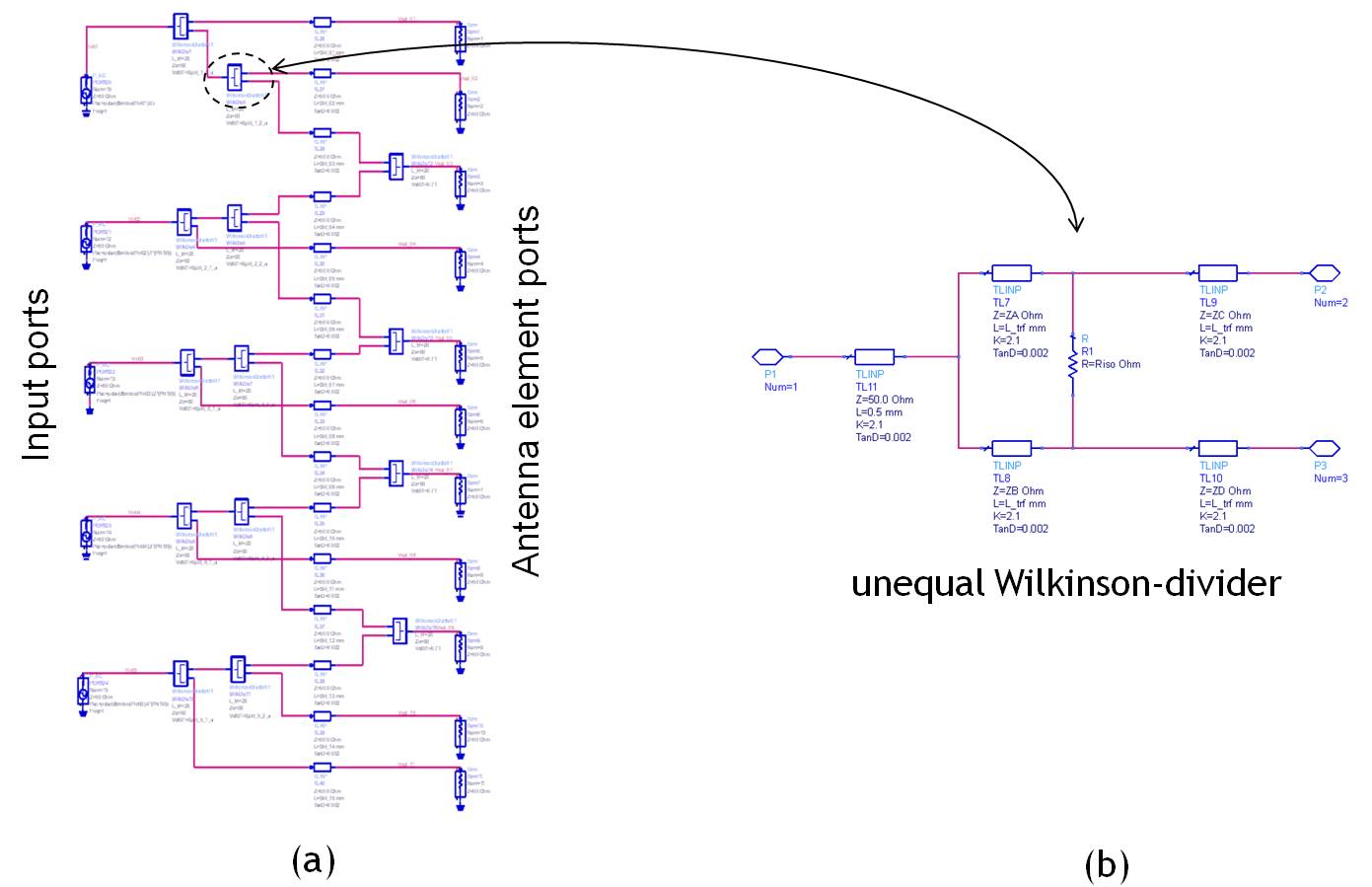}
\end{center}
\caption{\footnotesize{Macro-cell $11 \times 5$ RF beamformer instantiation: (a) RF beamformer factorized into two stages of WDs, $\bD_{w\,1}$ and $\bD_{w \, 2}$, followed by $\bP_{1}$ and one stage of directional couplers $\bR_{c\,1}$. (b) implementation details of an unbalanced WD.}}
\label{fig:10}
\end{figure*}

\subsection{Macro-Cell RF beamformer design with $\Ntrx=5$ and $\Nt=11$}
Fig. \ref{fig:10}(a) shows an $11 \times 5$ RFBN for a macro base station application. The implementation shows $\Ntrx=5$ voltage sources corresponding to the PA outputs and $\Nt = 11$ S-parameter-ports functioning as the input impedance of the antenna elements. This setup allows us to vary input voltages and phase shifts for different beamtilts and calculate the voltage and phase progressions at the RFBN output. We realise this design using micro strip lines on a typical dielectric substrate (dielectric constant = 3.48, loss tangent = 0.004), while considering the isolation resistance loss as well the micro strip loss.

The implementation follows from the design rules specified in Sections \ref{sec:3} - \ref{sec:4}. More specifically, the connections between transceiver chains and antenna elements, the power divider ratios as well as the phase shifts follow the algorithms and architectures specified in Sections \ref{sec:3} and \ref{sec:4}. We factorize the RFBN connections based on the \emph{Claim 1} and \emph{Claim 2} into multiple stages of Wilkinson dividers ($\bD_{w1}$ and $\bD_{w2}$), phase shift matrix ($\bP_{1}$) and DCs ($\bR_{c1}$). For simplicity, the current implementation contains only one stage of direction coupler $\bR_{c1}$. 


The five input signals are split successively into $N_{\bs}=15$ signal paths (using some unblanced splitters). For details on the implementation of unbalanced WDs, refer to Fig. \ref{fig:10}(b). Subsequently, micro-strip lines with varying line lengths are used to achieve the desired phase shifts  in order to beamform signals. Note that different incoming signal paths in $\bDfb$ pass through a varying number of dividers, and correspond to different lengths. To prevent insertion loss at  $\bRfb$, we include additional line lengths for the corresponding strip lines. Subsequently, the 15 signals are combined and coupled to 11 antennas through $\bRfb$ using DCs; for this implementation we use Wilkinson combiners.
 
\subsubsection{Beam pattern performance }
The RFBN has been initially designed for $\cR_{\theta} \in \{0^{\circ} \, - \, 15^{\circ}\}$. 
The main requirement in the optimization process is to minimize the combiner i.e. insertion loss before the antennas. As we aim to increase the range of beamtilts $\thetad$ for a given RFBN setup, we pay with poor beampattern performance and increased insertion loss. 
\begin{figure*}
\begin{center}
\includegraphics[width=0.75\textwidth]{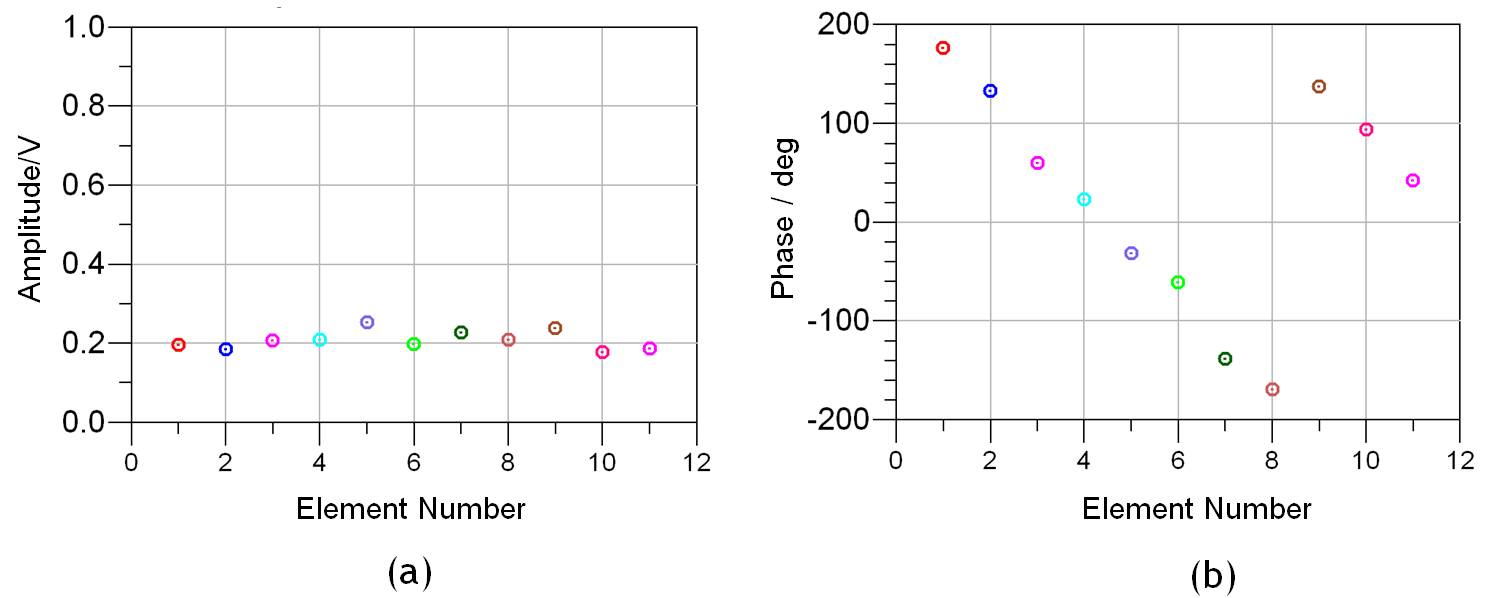}
\end{center}
\caption{\footnotesize{Macro-cell $11 \times 5$ RF beamformer performance: (a) Amplitude tapering and output voltage levels at  $\Nt=11$ antenna ports. (b) Phase tapering as we progress from antenna element 1 to antenna element 11. Note that the phase progression is linear to avoid spatial aliasing and near optimal beamforming.}}
\label{fig:11}
\end{figure*} 

\bef
\begin{center}
\includegraphics[width=\columnwidth]{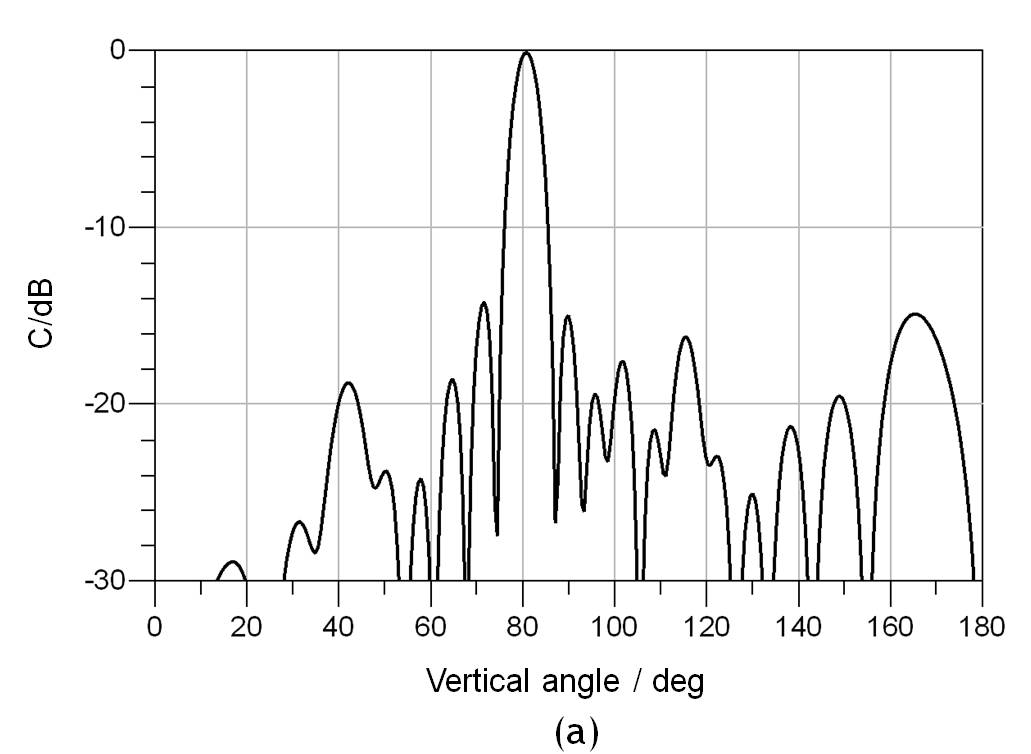}
\includegraphics[width=\columnwidth]{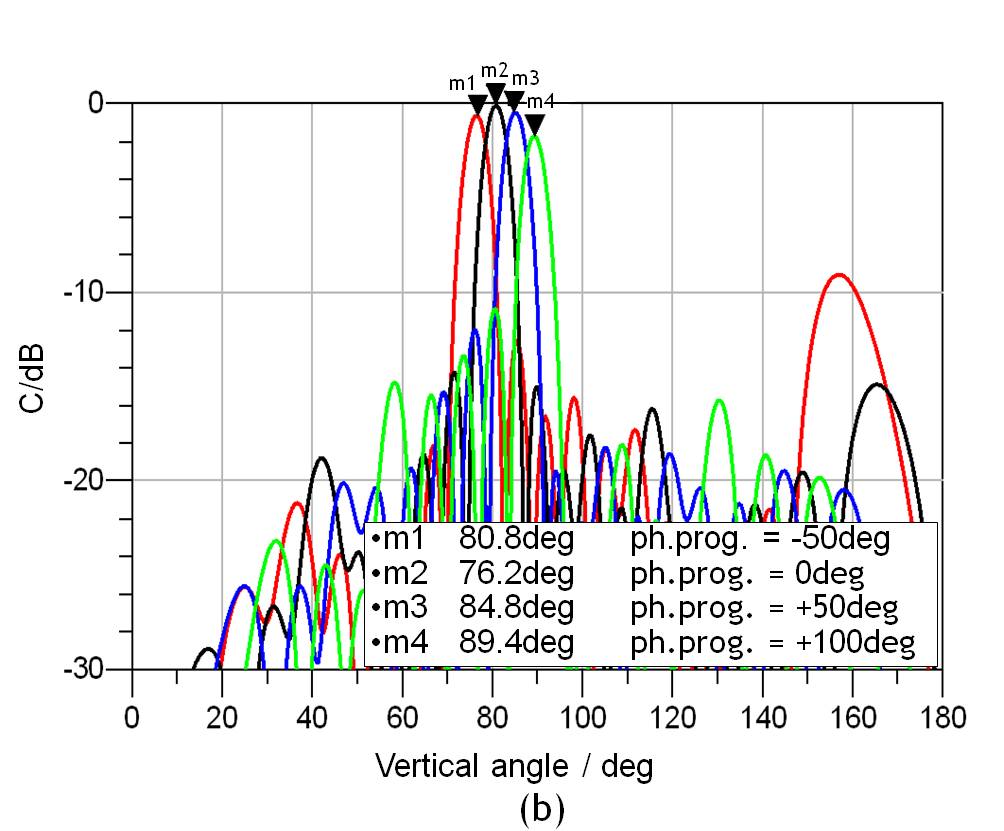}
\end{center}
\caption{\footnotesize{Macro-cell $11 \times 5$ RF beamformer beampattern: Beam-pattern performance of a $\Nt=11$ and $\Ntrx=5$ design provided with the same amplitude but with varying phase values at each RFBN output to achieve varying beamtilts. (a) Phase progression $0^{\circ}$ (b) phase progression $-50^{\circ}$, $+50^{\circ}$ and $+100^{\circ}$.}}
\label{fig:12}
\enf

As a sanity check for the RFBN design, we excite  $\Ntrx=5$ input signals with voltages of same magnitudes (with varying phase shifts). The beampattern quality for various beamtilts depends on the phase progression of the transmit signals within the RFBN. Fig. \ref{fig:11}(a) shows the voltage or amplitude levels at each antenna port for $\thetad=8^{\circ}$ and Fig. \ref{fig:11}(b) shows the phase progression of signals at each antenna element for $\thetad=8^{\circ}$. From Sec. \ref{sec:4} and Claim 2 we know of the advantages in designing a linear phase RFBN. We observe from Fig. \ref{fig:11}(b) that this property is preserved at each antenna output in our RFBN implementation. Linear phase progression of Fig. \ref{fig:11}(b) leads to a beampattern performance in Fig. \ref{fig:12}(a). 

Another sanity check of the RFBN design is to modify the phase shift of the signals excited at $\Ntrx=5$ inputs. 
As shown in Fig. \ref{fig:12}(b), if we apply an input phase progression of $-50^{\circ}$ from first input port to second input port, we observe a main lobe at $\thetad = -13.80^{\circ}$. Note that the SLLs of the main lobe (with marker m2) comply with 3GPP specs. Similarly, a phase progression of $+100^{\circ}$ results in a main lobe at $\thetad \approx0^{\circ}$ as in Fig. \ref{fig:12}(b). Note that these are simple-sanity checks on the design of RF beamformer. In practice, once we include the DBF algorithms, the overall performance improve as shown in Sec. \ref{sec:5}.

\subsubsection{Insertion loss performance}
Fig. \ref{fig:8a}(a) shows the average phase mismatch at the combiners of an  $11 \times 5$  setup for varying beamtilts $\thetad \in 0^{\circ}, \, \cdots, \, 30^{\circ}$. The insertion loss is proportional to the phase mismatch at the combiners (The rat-race couplers are balanced). Curve 1 shows the insertion loss for RFBN implemented using a lossy  Blass matrix \cite{Mosca:Blass} referred to in Sec. \ref{sec:4}. Curves 2 and 3 respectively show the performance of the proposed RFBN arrangement of Sec. \ref{sec:4}. From curve 3, we observe that as the beamtilt range increases, it becomes important to use hybrid couplers and compensate for the insertion loss.

The impact of the phase mismatch and network loss is shown in Fig. \ref{fig:8a}(b). This performance was obtained by subtracting the combined output power from combined input power for varying phase progressions at the input ports. The result shows loss in effective radiated power from the antenna array. For increasingly negative or positive phase increments from $\thetad=8^{\circ}$, the insertion loss starts to increase. These loss can be reduced by using two more stages of DCs.
 
\bef
\begin{center}
\includegraphics[width=\columnwidth]{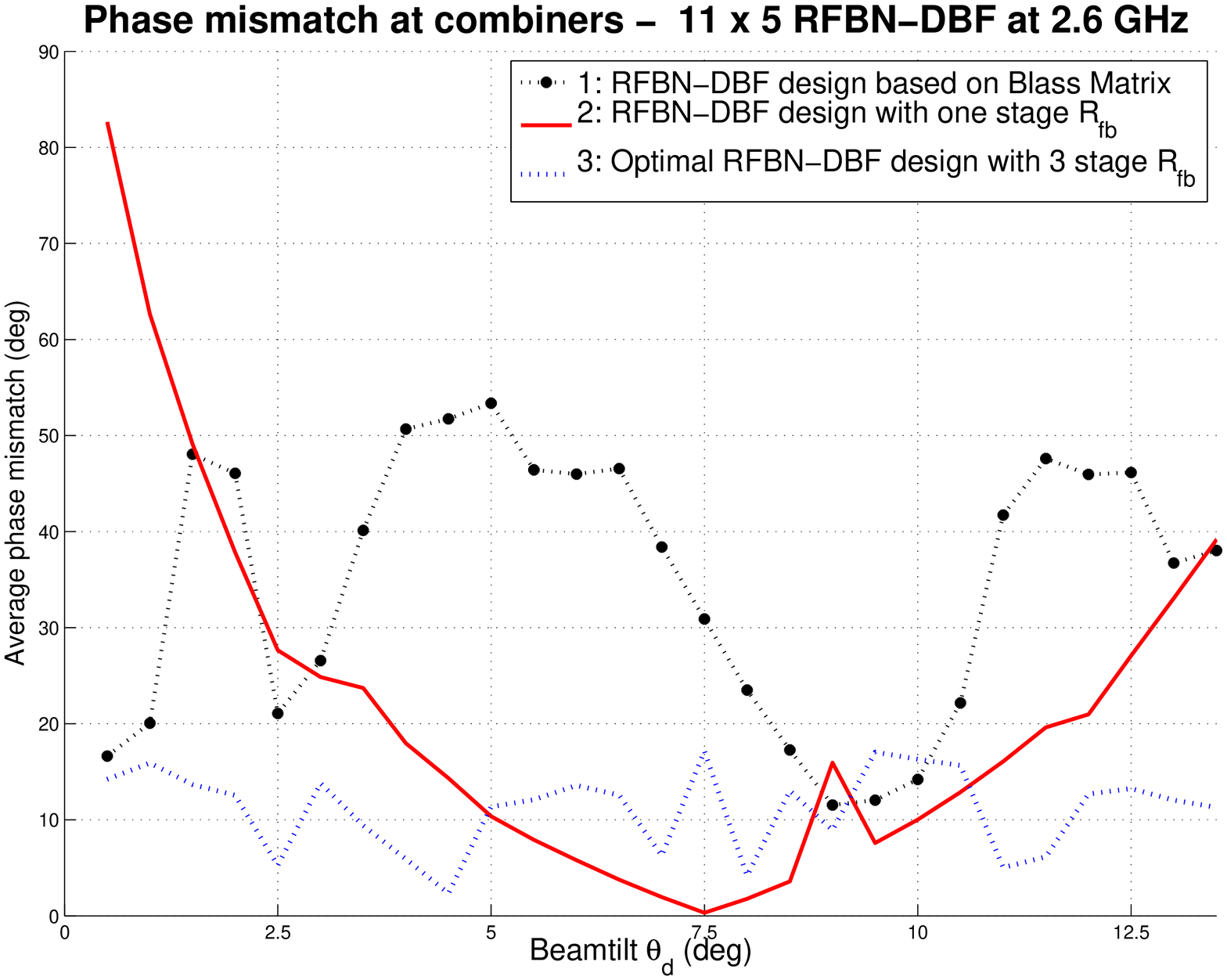}
\includegraphics[width=\columnwidth]{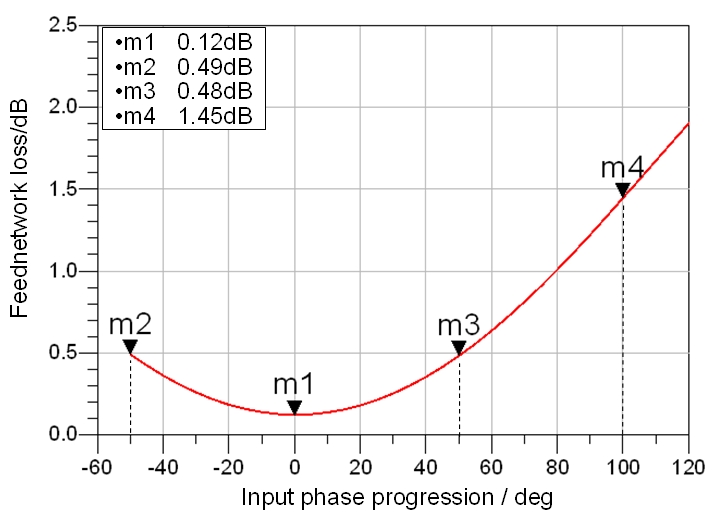}
\end{center}
\caption{\footnotesize{Insertion loss performance comparison: (a) Simulation performance of average phase mismatch between input signals at the last stage of $\bRfb$. Note that the overall insertion loss is directly proportional to the average phase mismatch at the last stage of $\bR_{fb}$. (b) Overall loss in radiated power due to insertion loss in the RFBN for configuration '2: RFBN-DBF design with one stage $\bRfb$'
}}
\label{fig:8a}
\enf

\subsection{small-cell RFBN design with $\Ntrx=3$ and $\Nt=6$}
Fig. \ref{fig:14} shows an $\Ntrx =3$ and $\Nt=6$ RFBN design for small-cell setup. The RFBN has been implemented using standard microstrip technology and the connections between RF chains and antenna elements follows Sec. \ref{sec:4b}. In the $\bDfb$ and $\bRfb$ stages, the ratios of all the employed power combiners and splitters vary from 0 dB to 12 dB. These components have been implemented either as unbalanced Wilkinson dividers (for power ratios up to 5 dBs) or alternatively as DCs  (when power ratios range from 5 dB to 12 dB). The phase shifts have been implemented using standard microstrip-based transmission lines, where the length of the line dictates the phase shift.

The small-cell RFBN is more complicated than the macro-RFBN design due  to orthogonal beampattern requirements. This example is composed of 5 discrete stages (2 stages of $\bDfb$, two stages of $\bRfb$ and one stage of $\bP_{1}$). The RFBN  inputs are generated by $\Ntrx=3$ transceivers $\bx=(x_1, x_2, x_3)$. The $1^{\textrm{st}}$ stage of RFBN is composed of three 1-to-3 Wilkinson power dividers, which split the signal of each transceiver output into three components and the output signals are phase-matched. The $2^{\textrm{nd}}$ stage of the RFBN is composed of nine 1-to-2 power dividers, leading to 18 ports after stage 2. The $3^{\textrm{rd}}$ stage of this RFBN $\bP_{1}$ is composed of eighteen static phase-shifting elements that match the phase of the $2^{\textrm{nd}}$ stage of the RFBN. 

In the $4^{\textrm{th}}$ stage of the RFBN,  six 2-to-1 power combiners are used to combine the amplitude and phase shift signal from the transceivers $x_2$ and $x_3$. 
The incoming signals at each combiner is phase matched and subsequently combined at the final stage of the combiners. Minimizing the phase mismatch for a given set of input signals should be one of the constraints as explained in Sec. \ref{sec:4}.  Finally, the last i.e., $5^{\textrm{th}}$ stage of this RFBN consists of 2-to-1 power combiners  coupling signals to $\Ntrx=6$ antennas.

\bef
\begin{center}
\includegraphics[width=0.8\columnwidth]{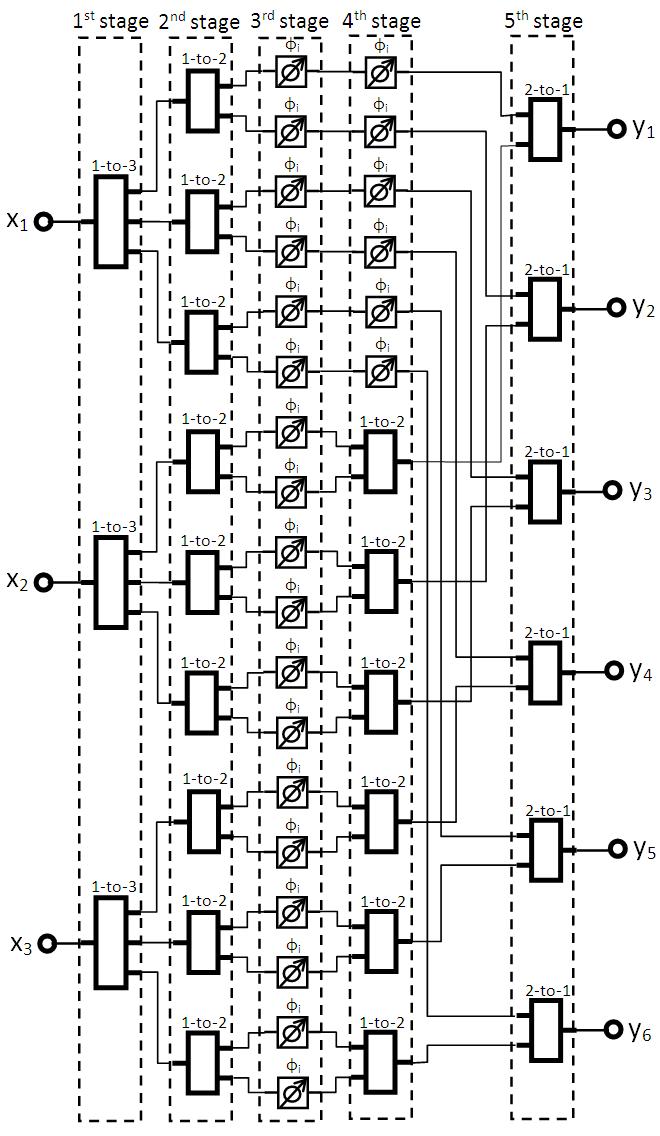}
\end{center}
\caption{\footnotesize{Small-cell RFBN instantiation with $\Nt=6, \, \Ntrx=3$  to provide distinct beams at $-30^{\circ}, \, 0^{\circ}, \, +30^{\circ}$}}
\label{fig:14}
\enf
\subsubsection{Beampattern performance}
\label{ssec:5_2}
A small-cell base station with RFBN antenna array typically beamforms and transmits the desired signal towards specific sector  $\thetad \in \{-30^{\circ}, \, \cdots, +30^{\circ}\}$.  In this case, the PAs typically radiate 0.5W of power. The base-station antennas are spaced $0.5 \lambda$ apart, and each antenna element has a 3-dB beamwidth of $110^{\circ}$. The amplitude tapering of the DBF weights connecting each PA is relaxed (with respect to macro-setup) to be in the range $0 - 3$ dB. The focus is to provide orthogonal beampatterns, while sacrificing on the insertion loss performance. 
\subsubsection{Effect of number of transceivers $\Ntrx$ and beamtilt range $\thetad$}
Fig. \ref{fig:15} shows the beampatterns for $\Ntrx=3,\, \Nt=6$, providing 3 sectors spaced $\thetad \in \{ -30^{\circ}, \, 0^{\circ}, \,  +30^{\circ} \}$. Note that the array response $\ba(\thetad), \, \thetad=\{-30^{\circ}, 0^{\circ} \}$ for each $\thetad$ is orthogonal to the other. We observe that it is possible to achieve 13 dB SLL suppression where all the PAs operating at a constant power. Curves 1, 3 and 5 in Fig. \ref{fig:15} compare the measured beampattern of the ADS implementation with the simulations results and we notice that the RFBN arrangement provides with 13 dB SLL.

\bef
\begin{center}
\includegraphics[width=\columnwidth]{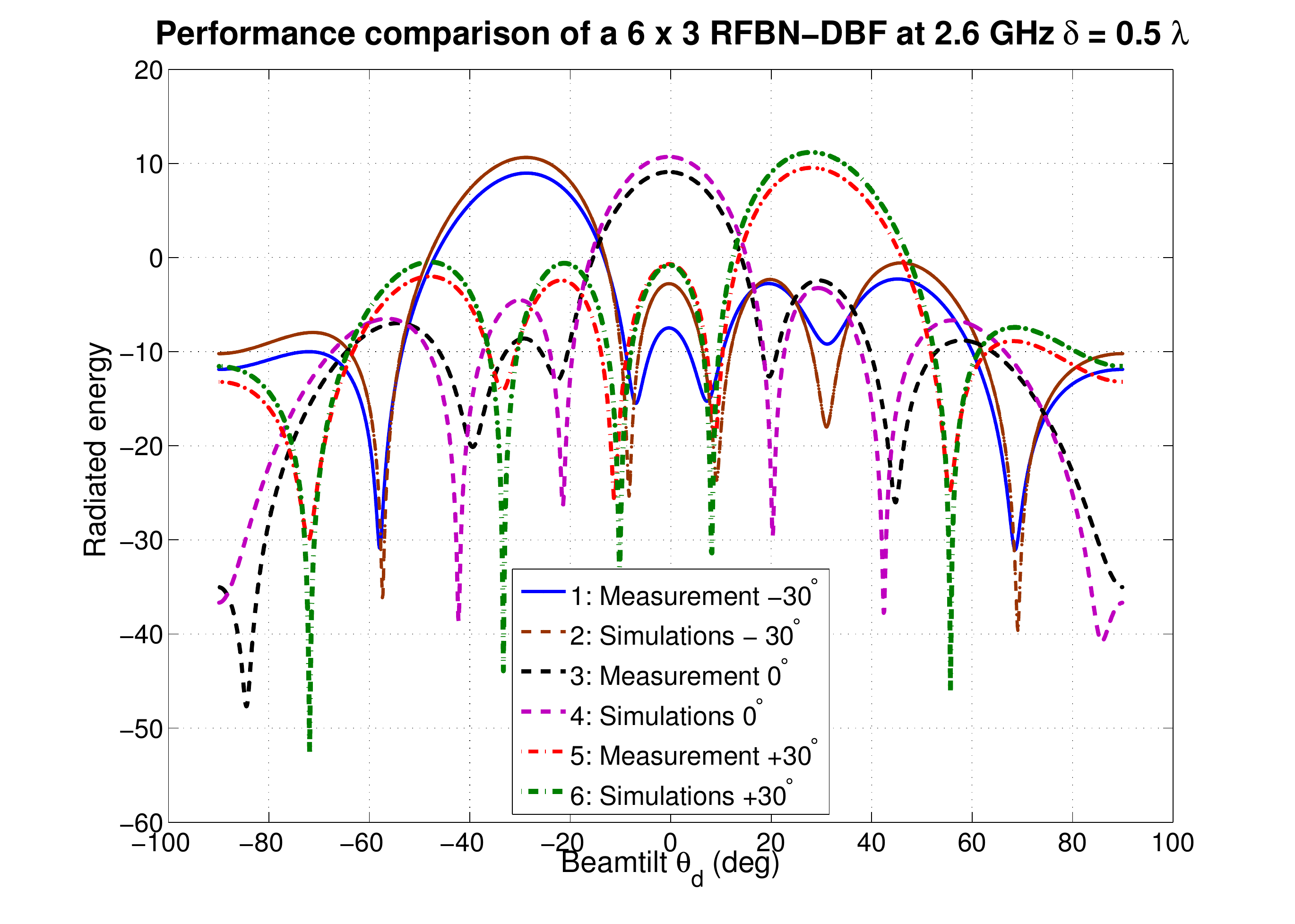}
\end{center}
\caption{\footnotesize{Small-cell joint RFBN-DBF to generate orthogonal beampatterns: $\Nt=6, \, \Ntrx=3$  to provide distinct main lobes at $-30^{\circ}, \, 0^{\circ}, \, +30^{\circ}$ with 13 dB sidelobe suppression.}}
\label{fig:15}
\enf


\section{Anechoic chamber RFBN measurements}
\label{sec:7}

To assess the capabilities of a macro-cell based RFBN, we implement an  $\Ntrx=5, \, \Nt=11$  RFBN detailed in Sec. \ref{sec:6} for 2.6 GHz and measure its beam-pattern performance in an anechoic chamber. 

\subsection{RFBN implementation}
The RFBN as shown in Fig. \ref{fig:17} is implemented with a  Roger 4350 substrate material. All power dividers, phase shifters, line crossings and combiners are optimized using using High-Frequency Structure Simulator (HFSS) \cite{Ansoft:hfss}, a commercial electromagnetic mode solver in order to meet the optimal performance of the individual components. In order to maximize the tilt range, a pre-tilt of 8 degrees was implemented,  which is the norm in a majority of cellular base station implementations. 
\bef
\begin{center}
\includegraphics[width=0.6\columnwidth]{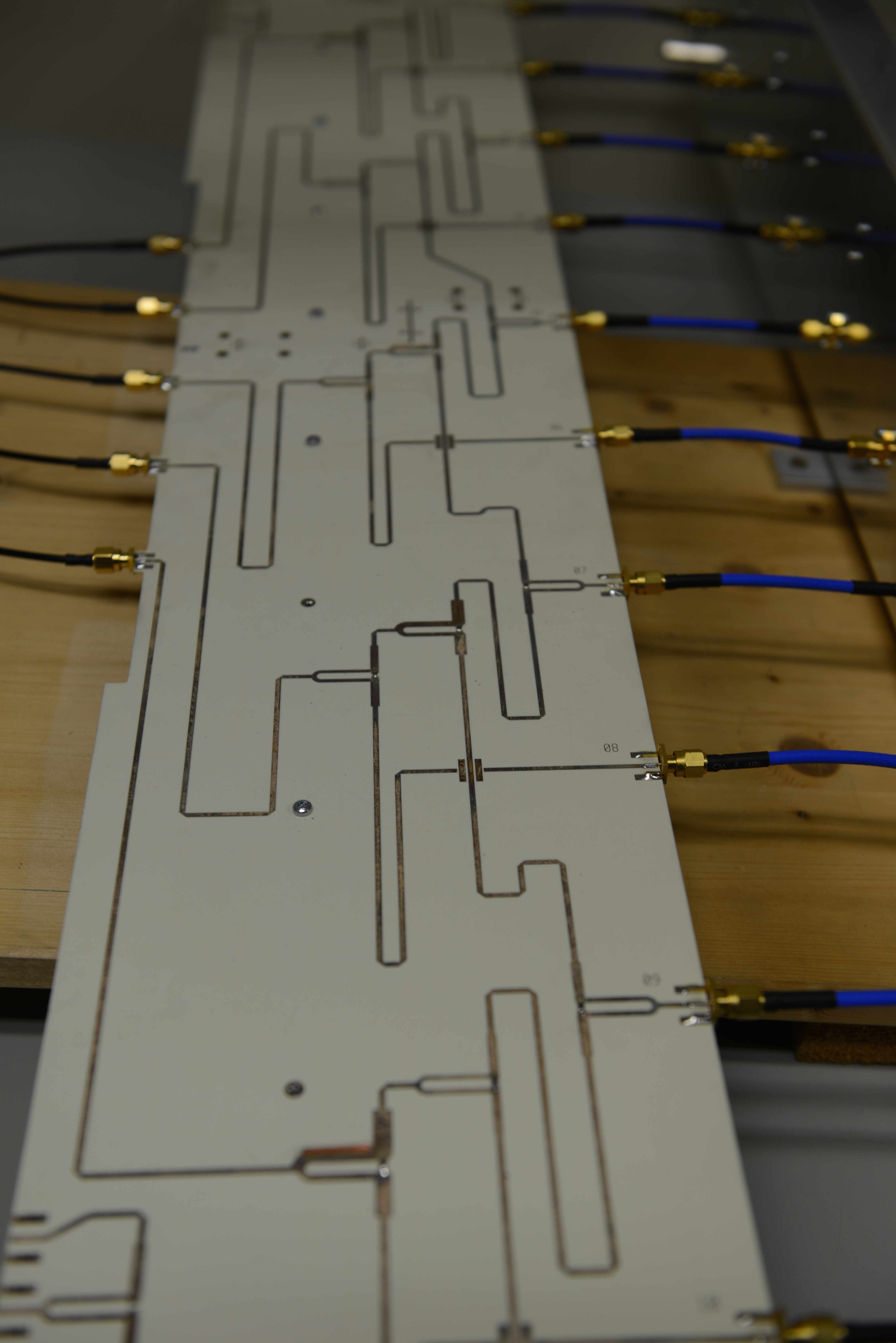}
\end{center}
\caption{\footnotesize{RFBN implementation with $\Ntrx=5$ input $\Nt=11$ output setup on Rogers 4350 substrate comprising of WDs, DCs and microstrip lines for phase shifts.}}
\label{fig:17}
\enf

\subsection{Measurement setup and calibration}
\noindent\subsubsection*{Measurement platform}
In order to test the performance of joint DBF-RFBN, we used multiple synchronized RF transmitters to generate beams from the device under test (DUT). The testing platform includes a hosting PC with graphic user interface (GUI) control panel, an RF transceiver board with multiple transmitters, an RFBN and an antenna array as shown in Fig. \ref{fig:15c}. In our system, a Xilinx Kintex-7 FPGA was used as the central DSP unit, and multiple complex digital phase shifters were implemented in the FPGA for tuning the phase of signal at each transmitter independently. Digitally-controlled step attenuators were used for tuning the amplitude of signal at each transmitter. 
\bef
\begin{center}
\includegraphics[width=0.8\columnwidth]{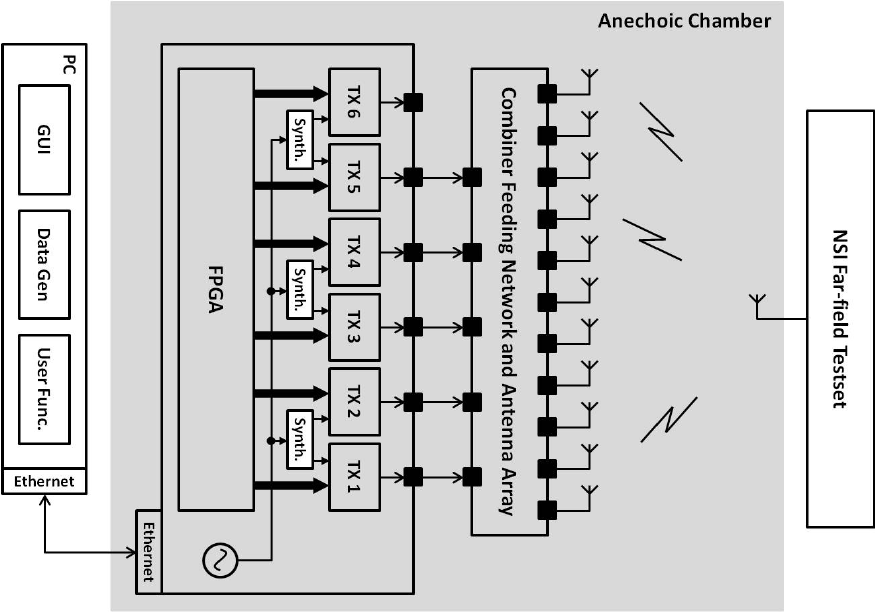}
\end{center}
\caption{\footnotesize{Anechoic chamber measurement setup}}
\label{fig:15c}
\enf

\noindent\subsubsection*{DBF and RF chain calibration}
Ideally, transmission path of each RF transmitter chain would be identical, in other words the phase and amplitude of multiple transmitter outputs would be exactly same if the source signal is the same. However, it is very difficult to layout and route multiple transmit chains on the PCB with equal length and same frequency response. It is also very difficult and expensive to guarantee that the connection cables are equal length and phase aligned. Additionally, the digital to RF transformation blocks, denoted as $\cR\cF\{.\}$ are not identical and the real physical system contains the following imperfections: (a) the phase lags of the signals from  digital baseband RFBN and antenna arrays are different and (b) the amplitude of the signals at the inputs of antenna arrays are not identical. 

In order to form desired beams with RFBN, phase and amplitude calibration are fundamentally required. The system calibration setup is illustrated in Fig. \ref{fig:15a}.

\bef
\begin{center}
\includegraphics[width=0.75\columnwidth]{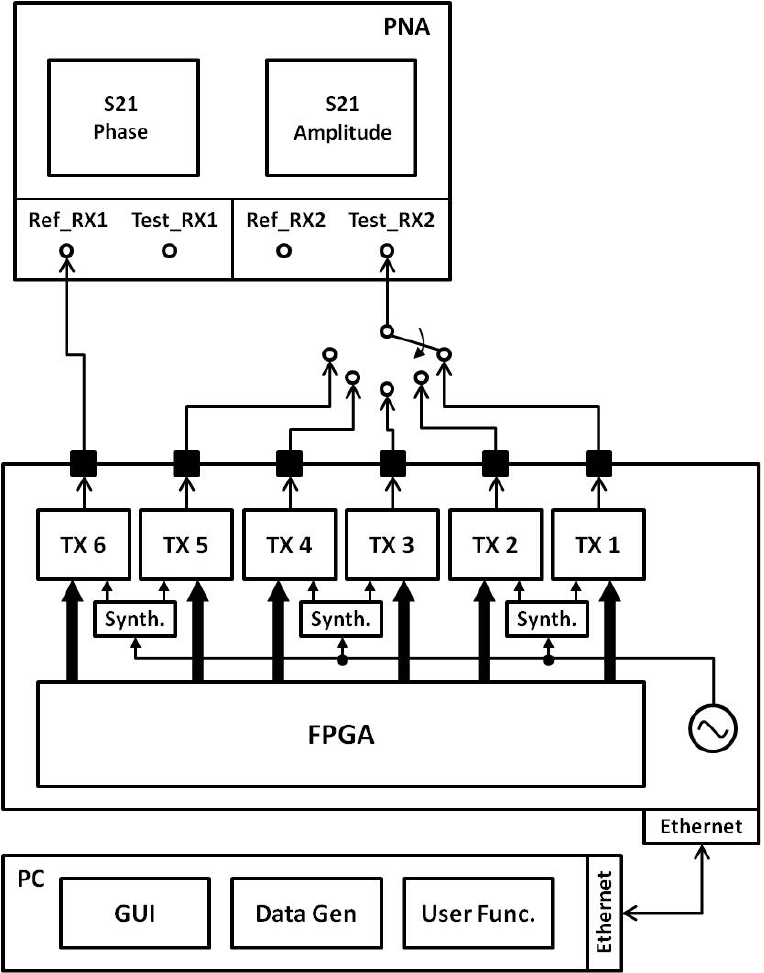}
\end{center}
\caption{\footnotesize{Calibration of active antenna array system: The signals at each RF chain is compared with reference signal (TX 6) for amplitude and phase offsets as well as phase drifts over a period of time.}}
\label{fig:15a}
\enf

\bef
\begin{center}
\includegraphics[width=0.8\columnwidth]{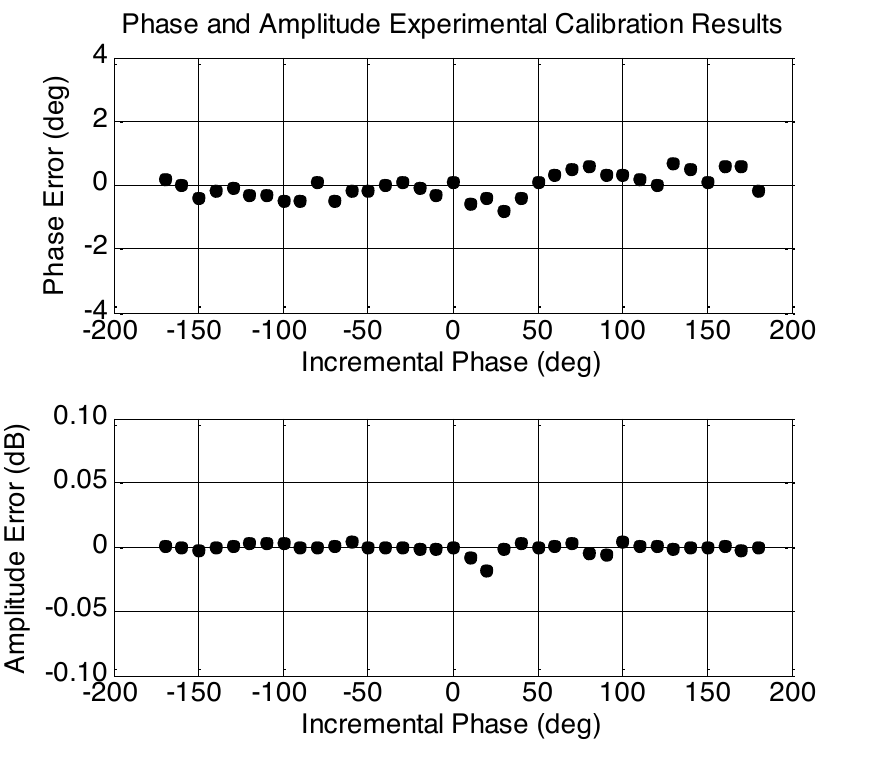}
\end{center}
\caption{\footnotesize{Average amplitude and phase error as we progress through the entire beamscanning range of network analyser.}}
\label{fig:15b}
\enf

A continuous wave (CW) signal is generated in the FPGA and upconverted to 2.6 GHz. We use a spare transmitter (TX 6) as reference and measure the relative phase difference and amplitude difference between the individual active transmitters TX 1 to TX 5 with  reference TX 6. Subsequently we add the corresponding phase and amplitude differences in the digital domain as offsets. After calibration, we sweep the phase from -180 degree to 180 degree and plot the difference between the input and output amplitude and phase values. 
Fig. \ref{fig:15b} plots the average phase and amplitude error as we sweep the DBF for all possible angles $(-\pi,\,\pi)$. The precision of the calibration setup is a function of this phase and amplitude error. We observe, that the average phase error is less than $\pm1^{\circ}$  and average amplitude error is less than $0.03$ dB. In other words, active transmitters are aligned in phase and amplitude after calibration.

\subsection{Performance evaluation antenna array setup}

In a typical macro-cell scenario, the base station is required to at least have a gain of 18 dBi along the main lobe, a 3-dB beam width of nearly $6^{\circ}$ and a SLL suppression in the order of at least -17 dB. In order to achieve vertical sectorization, the RFBN must account for the above set of gain and SLL requirements for the entire beamtilt range $\cR_{\theta} \in \{0^\circ, \, \cdots, \, 15^{\circ}\}$.  

The RF transceivers as well as the FPGA board operate at a data rate of 122.88 Msps. The device under test (DBF, transceivers, RFBN and antenna array setup shown in Fig. \ref{fig:15c}) is placed in the anechoic chamber and its beampattern performance is measured using an Agilent network analyzer. A common reference signal (with frequency 15.36 MHz) is used to synchronize the phase of FPGA output signals with that of the signals received at the network analyzer. 
Additionally, the beamtilts can be increased or decreased by applying the according amplitude and phase weights in the DBF resulting in a total tilt range from 0 to 16 degrees, while meeting the spatial mask and beamforming requirements required in a macro-cell base stations. Note that for each beamtilt, the power levels input to the PAs must be limited to a range of 0-1dB.

\bef
\begin{center}
\includegraphics[width=\columnwidth]{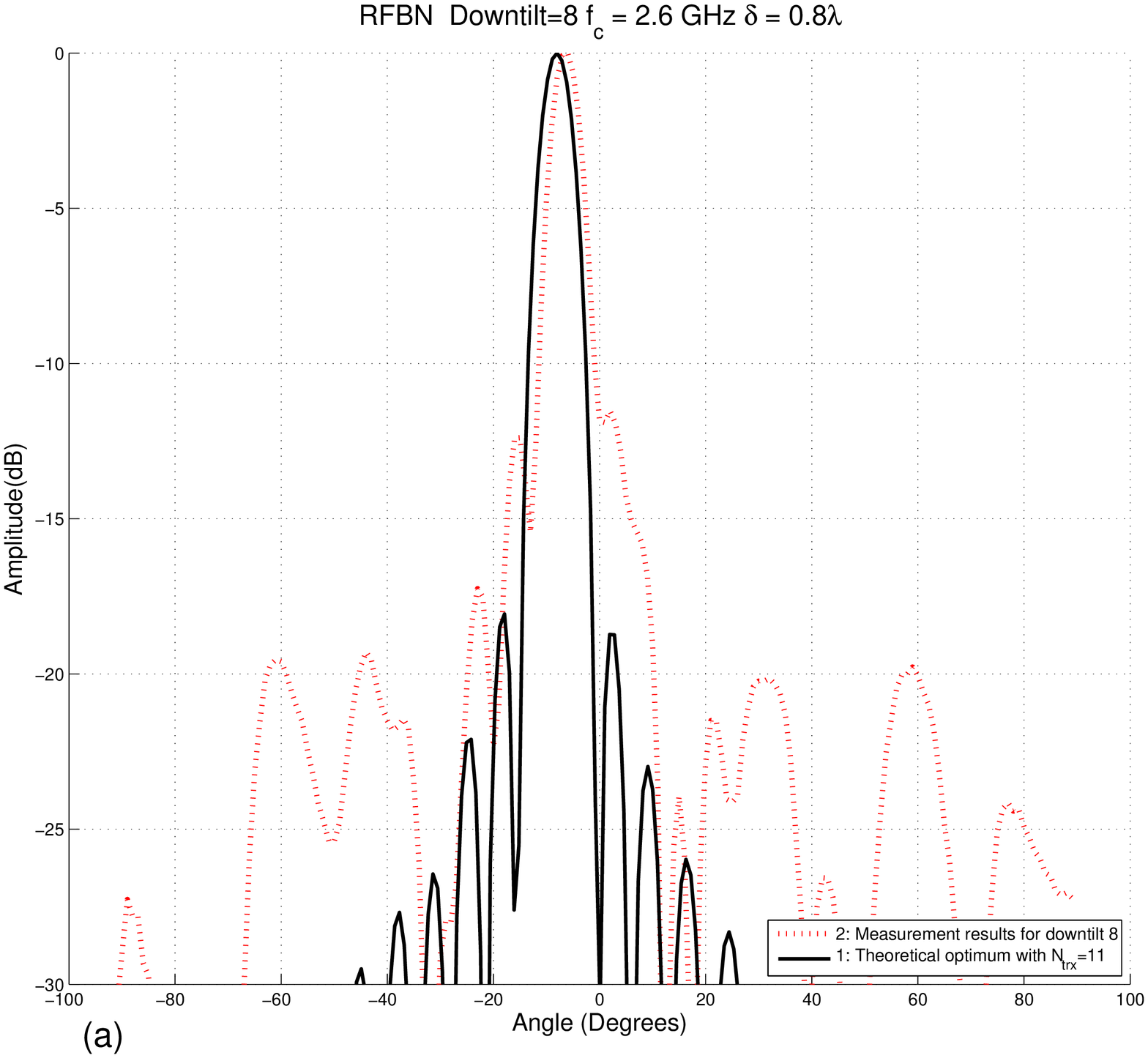}
\includegraphics[width=\columnwidth]{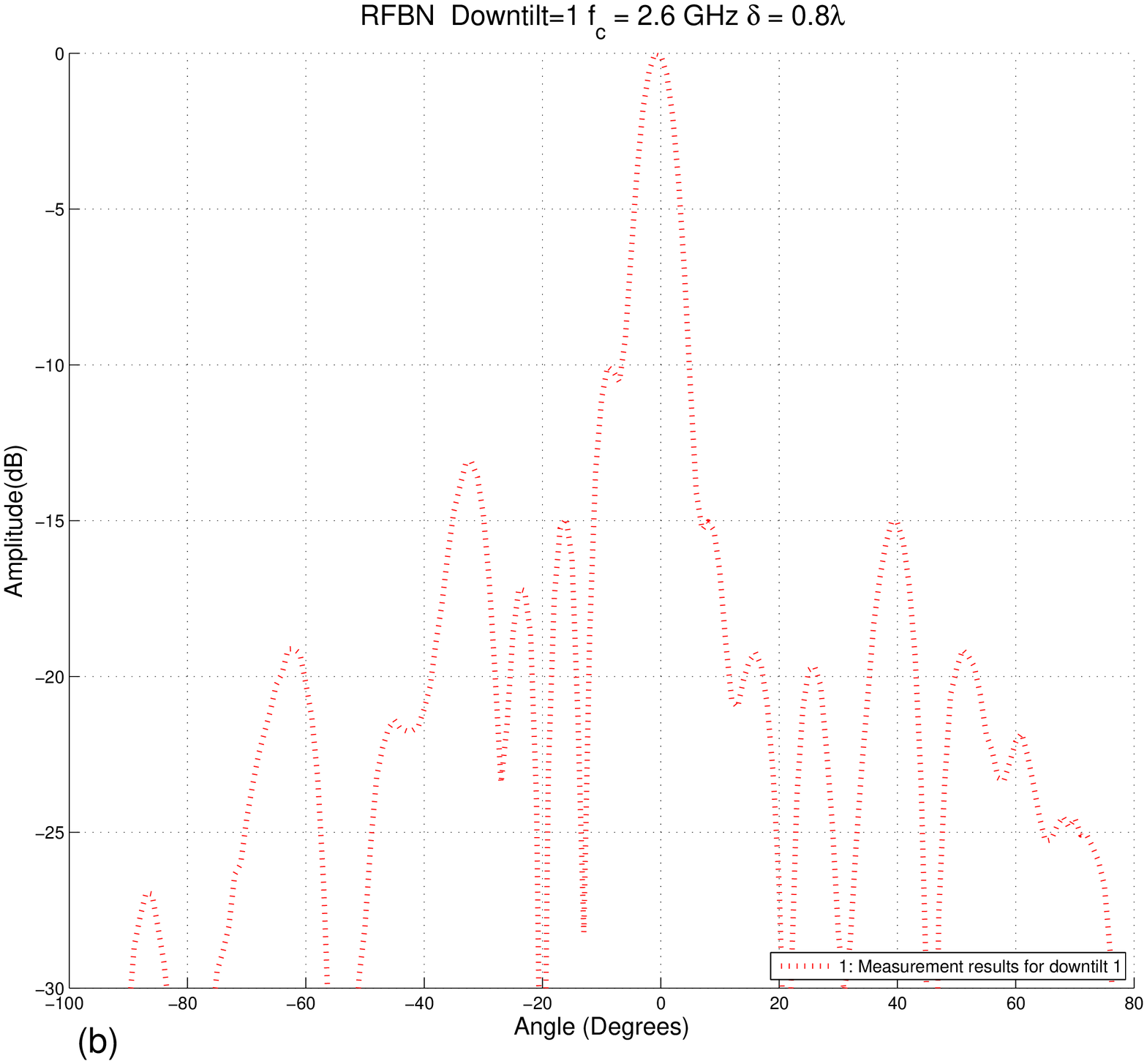}
\end{center}
\caption{\footnotesize{Macro-cell joint RFBN-DBF beampattern performance for $\Nt = 11$ and $\Ntrx=5$ at 2.6 GHz and $0.8\lambda$ spacing: (a) optimal setup for $\Ntrx=11$ transceivers and ideal PAs with infinite dynamic range for beam-tilt $8^{\circ}$ compared with $11 \times 5$ RFBN-DBF arrangement (b) RFBN-DBF setup with beam-tilt  $\thetad = 1^{\circ}$.}}
\label{fig:18}
\enf
\bef
\begin{center}
\includegraphics[width=\columnwidth]{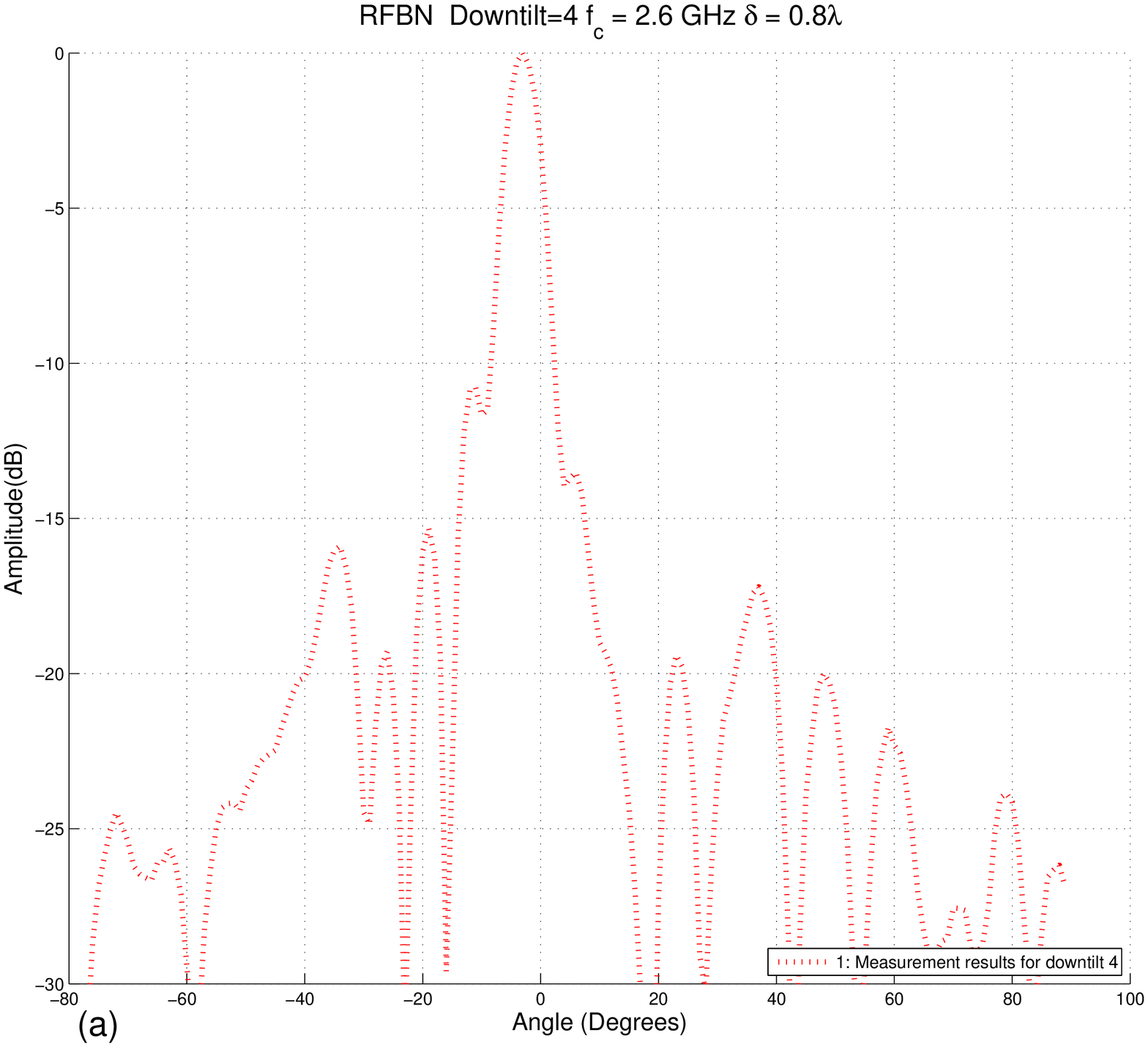}
\includegraphics[width=\columnwidth]{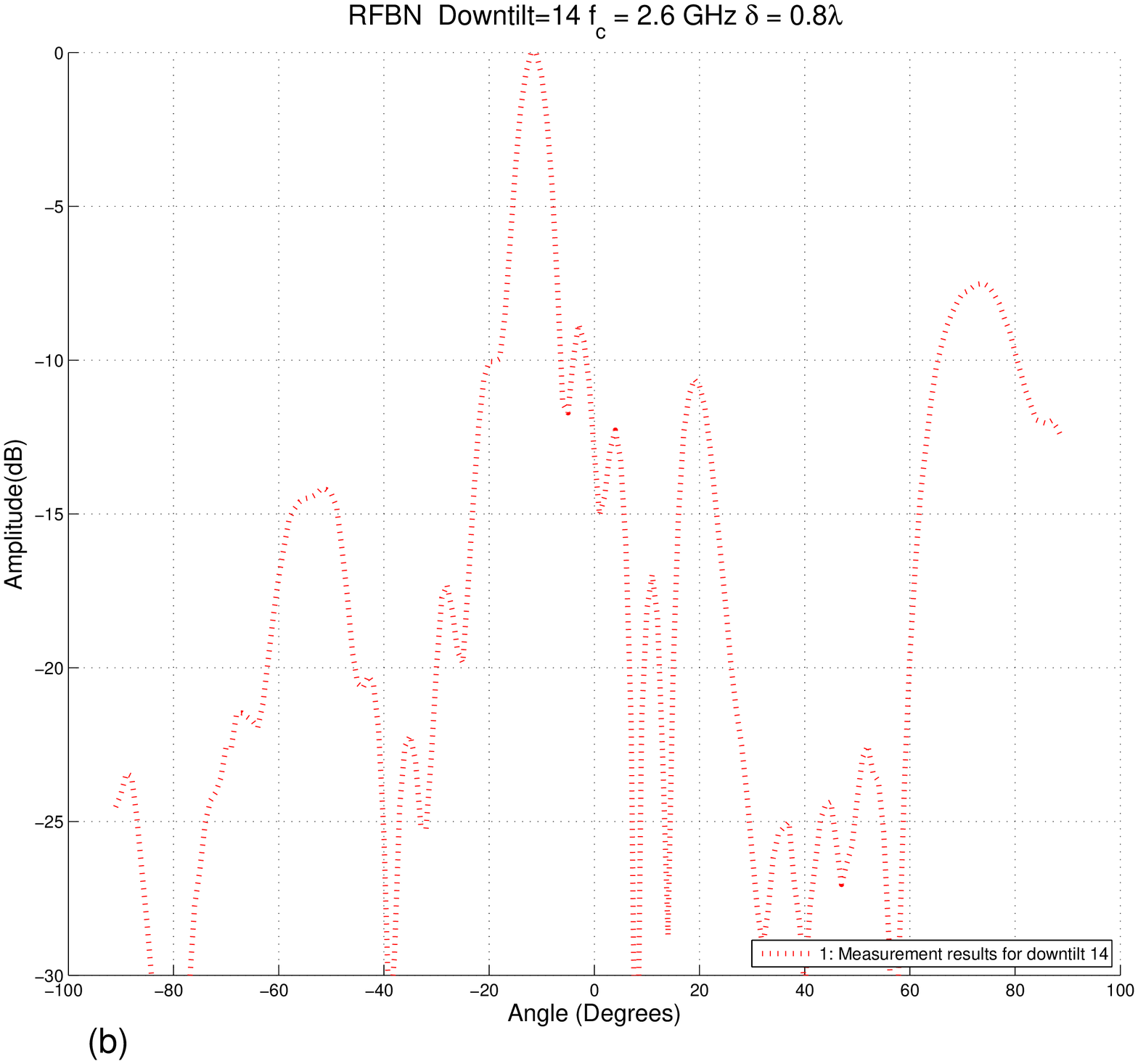}
\end{center}
\caption{\footnotesize{Macro-cell joint RFBN-DBF beampattern performance for $\Nt = 11$ and $\Ntrx=5$ at 2.6 GHz and $0.8\lambda$ spacing: (a)  RFBN-DBF performance for beam-tilt  $\thetad = 4^{\circ}$ (b) Beampattern performance measurements of macro-cell RFBN setup with $\Nt = 11$ and $\Ntrx=5$ at 2.6 GHz, $0.8\lambda$ spacing and  beamtilt = $14^{\circ}$.}}
\label{fig:19}
\enf
\bef
\begin{center}
\includegraphics[width=\columnwidth]{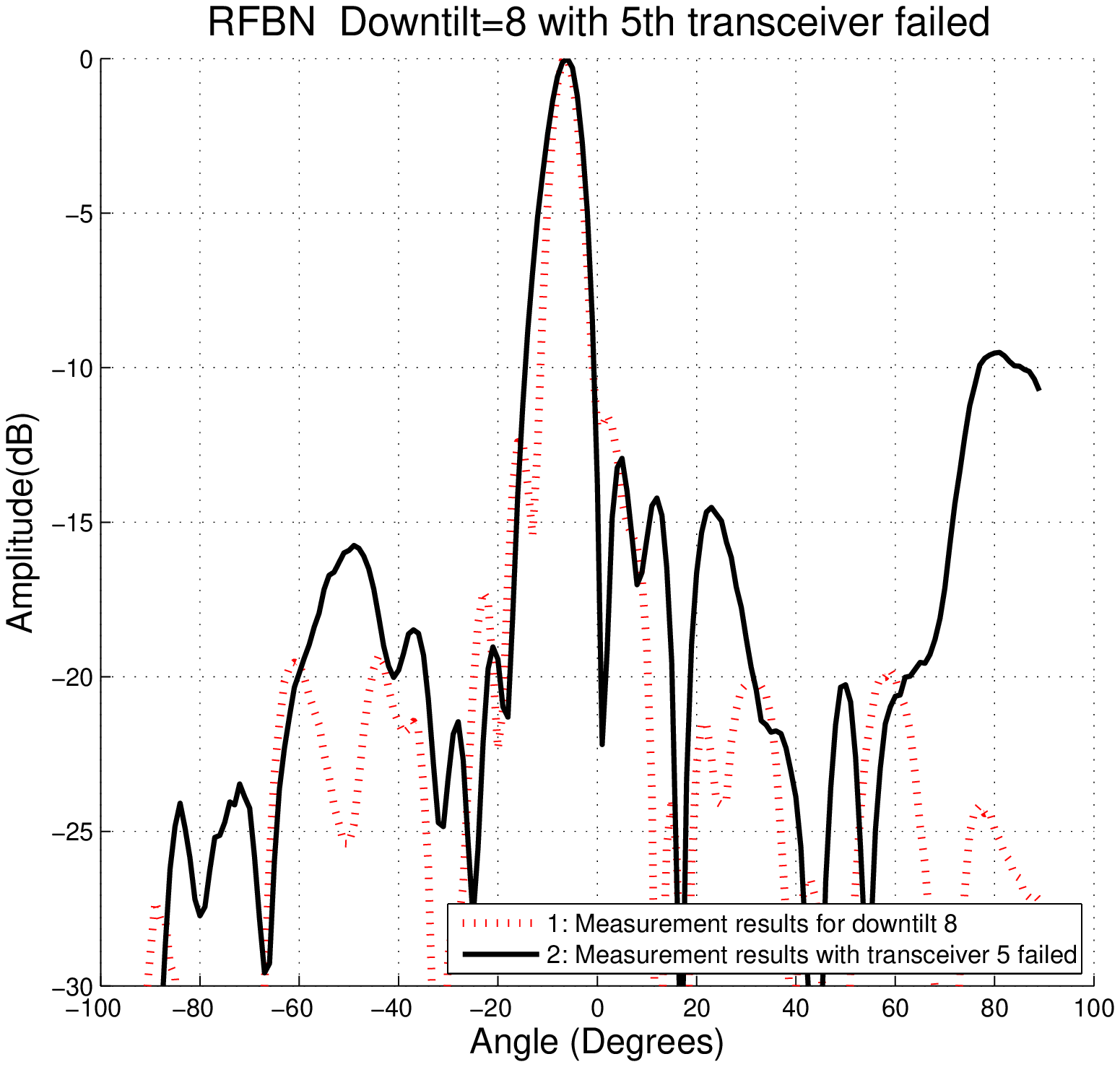}
\end{center}
\caption{\footnotesize{RFBN-DBF for beam-tilt  $\thetad = 8^{\circ}$ where the DBF is dynamically optimized if one of the transceiver fails}}
\label{fig:20}
\enf

Fig. \ref{fig:18}(a) shows measured beampattern of RFBN setup with network analyzer arrangement. The DBF gain/amplitude values are limited to be within a range of $0-1$ dB. Curve 1 shows the theoretical performance bounds when $\Ntrx=11$ transceivers  used without satisfying constraints [C3-C6]. Note that this is unrealistic in practical systems. Curve 2 shows anechoic chamber measurements for beamtilt $\thetad = 8 ^{\circ}$ and we observe that the main lobe is perfectly aligned and that the SLLs are around 18 dB below the main lobe.  

As the beamtilts are varied to an extreme case for beamtilt of $\thetad=1^{\circ}$, Fig. \ref{fig:18}(b) the main-lobe has a gain of 17 dBi, and the SLL is nearly 14 dB below the main lobe. Fig. \ref{fig:19}(a) shows the performance for $\thetad=4^{\circ}$ and shows that the SLL is 16 dB below main-lobe. As we push the DBF-RFBN setup to $\thetad =14^\circ$, the performance starts to degrade as shown in Fig. \ref{fig:19}(b), with SLL 10 dB below main lobe and grating lobes 8 dB below the main lobe. These are fundamental limitations due to spatial aliasing within the RFBN setup. 

One important property of the joint DBF-RFBN setup is its robustness. For example, if one element fails, the overall array can provide reasonably good performance with only $\Ntrx=4$ transceivers operating in degraded mode. Fig. \ref{fig:20} shows the measurement performance when TX 5 has failed, and only $\Ntrx=4$ are operational and connected to $\Nt=9$ antennas. Note that in a passive RFBN setup, the overall setup will be taken down and replaced. However, re-optimising the DBF for a degraded mode leads to a reasonably good performance. From curve 2, we observe that there is slight degradation in main-lobe energy and increase in the grating lobes due to spatial aliasing. However we must note that these are some of the worst case scenarios, since  array failure probability is around 2\%.

The same set of measurements can be repeated for a small-cell DBF-RFBN. 

\section{Conclusion}
\label{sec:8}

Existing RF beamforming networks are constructed in a mainly empirical way, which depending on the experience of the designer and the complexity of the required network often gives reasonable results. However, this process is very time consuming and does not guarantee the optimal solution in terms of beamforming performance, network complexity and minimize microwave loss.  A thorough understanding of the theoretical bounds as well as the microwave limitations will lead to the optimal solution enhancing the capacity and coverage of the communications system while operating at a reduced cost.  

For this reason, we have adopted a holistic approach and  proposed RFBN designs to reduce the number of transceivers while accounting the desirable features in next generation cellular base stations. Effectively,  we have showed how to determine the minimum number of transceiver elements in order to achieve a given set of access requirements and a presented a unified view on designing a hybrid beamforming network. Note that the two RFBN design requirements and designs differ a lot in terms of their system requirements: 
\bds
\item RFBN for a small-cell base station, generating three static beams with a rather broad beam width in the horizontal direction and tilt range of $(-30^{\circ}, 0^{\circ}, +30^{\circ})$. In this design the focus was on achieving a set of orthogonal beams while maintaining a side lobe suppression of at least 10 dB.
\item RFBN for a macro-cell base station antenna with a sharp and narrow vertical beam providing a continuous beamtilt range anywhere between $\thetad \in \{0^{\circ}, \cdots, \, 15^{\circ}\}$, while maintaining a tight set of spectral mask, sidelobe level and insertion loss requirements. We verify its performance in anechoic chamber.
\eds
Despite these two very different sets of requirements in both applications, the two derived RFBNs prove benefits in designing through a signal theoretic approach. Especially in the case of a macro-cell base station antenna we could show that the overall loss is kept to a minimum, which is essential for applications where the amount of radiated power easily reaches 100W. Note that the overall loss in the RFBN leads to reduced radiated power as well as additional problems in thermal management. Some of the future research directions are as follows:
\bds
\item In small-cell networks, combiner loss is not critical in terms of thermal management when radiated power levels $\leq 5$W. For example, an insertion loss of 1dB results in approximately 0.6W through heat dissipation. However they affect the communication range and receiver sensitivity. For this reason, future small-cell networks must consider insertion loss minimization in addition to wide beamtilt ranges.
\item Another aspect of our future work is the expansion to next generation communication systems such as millimeter wave \cite{Roh:mmwave} and large scale antenna arrays \cite{Marzetta:large_scale_mimo}. It is reasonably clear that the additional cost incurred due to increased number of transceivers  for large scale arrays will limit their widespread use. Cost effective construction of large RFBNs and advanced 2-dimensional beam steering methods will ensure that the theoretical benefits of massive MIMO and millimeter wave communication systems are realized in practice.
\eds

\bibliographystyle{plain}
\bibliography{mybibliography}

\end{document}